\documentclass[11pt]{article}

\usepackage{amsmath,amssymb,esint}
\usepackage[scale=0.8]{geometry}
\usepackage{cite}
\usepackage[normalem]{ulem}
\usepackage{graphicx}
\usepackage{subcaption}
\usepackage{float}
\usepackage{bbold}
\usepackage{enumitem}

\usepackage{caption}
\usepackage{amssymb}
\usepackage{amsthm}
\theoremstyle{plain}
\numberwithin{equation}{section}
\usepackage{multirow}
\usepackage{wasysym}
\usepackage{dynkin-diagrams}
\usepackage{arydshln}
\usepackage{mathtools}

\usepackage{esvect}
\usepackage{physics}
\usepackage{tensor}
\usepackage{bm}
\usepackage{color}
\usepackage{hyperref}
\usepackage{slashed}
\usepackage{blkarray}

\newcommand{\cZ}{\mathcal{Z}}

\newcommand{\cB}{\mathcal{B}}
\newcommand{\cF}{\mathcal{F}}

\newcommand{\cV}{{\cal V}}
\newcommand{\cW}{{\cal W}}
\newcommand{\cK}{{\cal K}}

\newcommand{\tkap}{\tilde{\kappa}}
\usepackage{listings}

\usepackage{tikz}
\usetikzlibrary{arrows,matrix,positioning,shapes,arrows}
\usetikzlibrary{shapes.geometric, arrows, calc, intersections}
\usetikzlibrary{positioning}

\usepackage{youngtab}
\usepackage{changepage}

\newtheorem{thm}{Theorem}[section]

\newcommand*\xbar[1]{%
  \hbox{%
    \vbox{%
      \hrule height 0.5pt 
      \kern0.4ex
      \hbox{%
        \kern-0.25em
        \ensuremath{#1}%
        \kern-0.1em
      }%
    }%
  }%
} 

\definecolor{oxfordblue}{rgb}{0.0, 0.13, 0.28}

\usepackage[most]{tcolorbox}
\makeatletter
\def\@endtheorem{\endtrivlist\@endpefalse\@afterheading}
\makeatother

\begin{document}
\phantom{}
\vspace{-.2cm}

\begin{center}
{\LARGE {\bf {\Large\bfseries Heterotic Flux Vacua with a Small Superpotential}}}\\[24pt]
{\bf{Evgeny I. Buchbinder$^{a,}$\footnote{evgeny.buchbinder@uwa.edu.au},
Andrei Constantin$^{b, c,}$\footnote{andrei.constantin@physics.ox.ac.uk}, Lucas T.\ Y.\ Leung$^{c,}$\footnote{lucas.leung@physics.ox.ac.uk},\\ Andre Lukas$^{c,}$\footnote{lukas@physics.ox.ac.uk}, Burt Ovrut$^{d,}$\footnote{ovrut@physics.upenn.edu}}}
\bigskip\\[0pt]
\vspace{0.23cm}
${}^a$ {\it Department of Physics M013, The University of Western Australia,\\ 35 Stirling Highway, Perth WA 6009, Australia
}
\vspace{12pt}

${}^b$ {\it 
School of Mathematics, University of Birmingham, Watson Building,\\ Edgbaston, Birmingham B15 2TT, United Kingdom
}
\vspace{12pt}

${}^c$ {\it 
Rudolf Peierls Centre for Theoretical Physics, University of Oxford\\
Parks Road, Oxford OX1 3PU, United Kingdom
}
\vspace{12pt}

${}^d$ {\it 
Department of Physics, University of Pennsylvania, Philadelphia, PA 19104, USA
}
\end{center}
\vspace{.5cm}

\abstract{
\noindent 
We study heterotic Calabi--Yau compactifications with NSNS three-form flux in view of moduli stabilisation and investigate whether the value 
$|W_0|$ of the flux superpotential evaluated at supersymmetric minima can be small. Unlike in type~IIB string theory, heterotic compactifications lack a no-scale structure, so that a non-vanishing flux superpotential generically induces a tree-level scalar potential for all moduli. Controlled moduli stabilisation therefore requires the flux superpotential to be sufficiently small in order to compete with non-perturbative effects. Working within a four-dimensional effective field theory and exploiting the special geometry of Calabi--Yau complex structure moduli spaces, we analyse the complex structure F-term equations and derive two no-go theorems: (1) supersymmetric vacua with vanishing $|W_0|$, which would lead to a vanishing tree-level scalar potential as in type~IIB, occur only at singular loci in moduli space, and (2) no supersymmetric vacua with small $|W_0|$ exist in the large complex structure limit. Motivated by these results, we analyse explicit models away from the large complex structure regime using exact period expressions and identify one- and two-parameter examples in which suitable flux choices allow for values of $|W_0|$ of order unity, with some one-parameter models admitting values moderately below unity.
Our findings show that small heterotic flux superpotentials are highly constrained, with the values of $|W_0|$ found in the examples studied being only marginally compatible with moduli stabilisation.

\newpage

\tableofcontents

\section{Introduction} \label{sec:intro}
Moduli stabilisation, the problem of generating a scalar potential that fixes the vacuum expectation values of moduli fields, remains one of the central open challenges in string theory~\cite{Denef:2004ze,Grana:2005jc,Ibanez:2012zz,McAllister:2023vgy}. Resolving this issue is essential not only to avoid phenomenologically unacceptable long-range ``fifth forces'', but also because moduli VEVs determine the parameters of the low-energy theory as well as the scale and structure of supersymmetry breaking. A widely studied strategy is to combine flux-induced superpotentials with non-perturbative effects, such as instantons or gaugino condensation, in order to generate non-trivial minima of the scalar potential. For this approach to be viable, the tree-level flux contribution must be sufficiently small so that it can compete with the exponentially suppressed non-perturbative terms and other subleading corrections. The use of fluxes, gaugino condensation, and instanton effects for moduli stabilisation has a long history. In the following, we emphasise developments in the heterotic context, which is the focus of this work; see, for example, Refs.~\cite{Dine:1985rz,Derendinger:1985kk,Dine:1986zy,Dine:1987bq,Font:1990nt,Casas:1990qi,Ferrara:1990ei,Nilles:1990jv,Witten:1996bn,Lukas:1997fg,Witten:1999eg,Gukov:1999ya,Giddings:2001yu,Kachru:2003aw,deCarlos:2005kh,Balasubramanian:2005zx,Conlon:2005ki,Buchbinder:2002ic,Buchbinder:2002pr,Buchbinder:2003pi,Gurrieri:2004dt,Lukas:1997rb,Gray:2007qy,Braun:2007xh,Braun:2007tp,Braun:2007vy,Anderson:2011cza,Cicoli:2013rwa,delaOssa:2015maa,Buchbinder:2016rmw,Buchbinder:2017azb,Buchbinder:2018hns,Buchbinder:2019hyb,Buchbinder:2019eal,Demirtas:2019sip,Demirtas:2021nlu,Deffayet:2023bpo,Deffayet:2024hug}.
\\[2mm]
In many type~IIB compactifications this requirement is alleviated by the presence of a no-scale cancellation structure at leading order~\cite{Giddings:2001yu,Grana:2005jc,McAllister:2023vgy}. This is a consequence of the IIB flux superpotential depending on the complex structure moduli and the axio-dilaton, but not on the K\"ahler moduli. Fluxes therefore stabilise the complex structure moduli and the axio-dilaton at tree level, while the K\"ahler moduli remain flat with a vanishing scalar potential along these directions. This separation allows the K\"ahler moduli to be stabilised by subleading non-perturbative or $\alpha'$ corrections, even when the flux superpotential is not small~\cite{Kachru:2003aw,Balasubramanian:2005zx,Conlon:2005ki}.\\[2mm]
By contrast, heterotic Calabi--Yau compactifications do not exhibit a no-scale cancellation structure. This is because the heterotic flux superpotential depends only on the complex structure moduli and does not involve the axio-dilaton, which is therefore not fixed by the F-term equations at tree level. To discuss the implications in more detail let us introduce the quantities
\begin{equation}\label{U0}
W_0=\langle\hat{W}\rangle_{F_a=0}\;,\qquad
U_0=\langle e^{\hat{K}/2}\hat{W}\rangle_{F_a=0}\;,
\end{equation}
where $\hat K$ and $\hat W$ denote the K\"ahler potential and superpotential expressed in affine complex structure moduli coordinates and  $F_a=0$ refers to the F-term equations for the complex structure moduli. The quantity $W_0$ corresponds to the vacuum value of the superpotential and is not K\"ahler invariant, while $U_0$ provides a K\"ahler invariant measure of its physical scale. The magnitude $|W_0|$ can be compared to that of the non-perturbative superpotential, whereas $|U_0|$ controls the overall scale entering the scalar potential.\\[2mm] 
Due to the absence of a no-scale structure, a non-vanishing flux superpotential induces a tree-level contribution to the scalar potential proportional to $|U_0|^2$.
Consequently, non-perturbative effects can only stabilise the remaining moduli if their contributions to the superpotential are comparable to the flux term. Since non-perturbative effects are exponentially suppressed, this will certainly not work if $|W_0|\gg 1$. Achieving sufficiently small values of $|W_0|$ is therefore a structural necessity for controlled moduli stabilisation in heterotic models.\\[2mm]
The challenge is further compounded by the restricted form of the heterotic flux superpotential. Unlike in type~IIB string theory, where the flux superpotential depends on the complex structure moduli, the axio-dilaton, and both Neveu--Schwarz Neveu--Schwarz (NSNS) and Ramond--Ramond~(RR) fluxes, the heterotic flux superpotential depends only on the complex structure moduli and NSNS three-form flux. As a result, the available flux choices are significantly more constrained, and it is an open question whether sufficiently small values of $|W_0|$ can be realised. Answering this question is crucial for heterotic moduli stabilisation and will be the main purpose of this paper.\\[2mm]
The paper is organised as follows. In Section~\ref{sec:background} we set up the notation and state the main question more precisely. In Section~\ref{sec:no-gos} we prove two relevant no-go theorems, one excluding viable solutions to $d\hat{W}=\hat{W}=0$ for the heterotic flux superpotential $\hat{W}$, the other excluding vacua with small $|W_0|$ in the large complex structure limit. Motivated by these no-go theorems, we devote the remainder of the paper to the study of explicit examples, analysing flux potentials away from the large complex structure limit using exact expressions for the period integrals. In Section~\ref{sec:one-param} this is done for a number of one parameter models and in Section~\ref{sec:two-param} for a two-parameter model. We conclude in Section~\ref{sec:conclusion}. The appendices contain a detailed proof of the second no-go theorem and the period expressions used in the examples. 

\section{Heterotic CY compactifications with NSNS flux}
\label{sec:background}

To set the scene and fix the notation, we start by describing the relevant parts of the four-dimensional $N=1$ supergravity obtained from heterotic compactifications on a CY three-fold $X$ with NSNS three-form flux $H$.
The gravitational moduli include the axio-dilaton $S$ with $s={\rm Re}(S)$ and the K\"ahler moduli $T^i$ with $t^i={\rm Re}(T^i)$, where $i= 1,\ldots ,h^{1,1}(X)$. Our focus will be on the complex structure moduli space with homogeneous coordinates $\cZ^A$, where $A=0,1,\ldots,h^{2,1}(X)$ and their affine counterparts $Z^a=\cZ^a/\cZ^0$, where $a=1,\ldots ,h^{2,1}(X)$, on the patch with $\cZ^0\neq 0$.\\[2mm]
{\bfseries Structure of supergravity.} Within four-dimensional $\mathcal N=1$ supergravity, the F-term scalar potential is determined by the superpotential $\cW$ and K\"ahler potential $\cK$ and is given by
\begin{equation}\label{V}
V = e^{\cK}\left(\cK^{I\bar J} D_I\cW\, D_{\bar J}\overline{\cW} - 3|\cW|^2\right),
\qquad
D_I\cW = \cW_I + \cK_I\,\cW \, ,
\end{equation}
where $\cW_I=\partial \cW/\partial\Phi^I$, $\cK_I=\partial \cK/\partial \Phi^I$, and $\Phi^I$ denotes all chiral superfields. In our case, the superpotential has the general structure
\begin{equation}\label{Wfull}
 \cW = W(\cZ) + W_{\rm np}(\cZ,S,T,\ldots)\; .
\end{equation}
In this paper we set $W_{\rm np}=0$ and focus on the flux superpotential $W(\cZ)$. To write down its explicit form, we briefly recall the special geometry of complex structure moduli space encoded in the prepotential $\cF(\cZ)$, a holomorphic homogeneous function of degree two, and its derivatives $\cF_A=\partial \cF/\partial\cZ^A$. The holomorphic $(3,0)$-form $\Omega\in H^3(X,\mathbb C)$ is expanded as
\begin{equation}
 \Omega = \cZ^A \alpha_A - \cF_A \beta^A
 = C^T \eta\, \Pi \;,\qquad
 C=\begin{pmatrix}\alpha_A\\ \beta^A\end{pmatrix},\quad
 \Pi=\begin{pmatrix}\cF_A\\ \cZ^A\end{pmatrix},\quad
 \eta = \begin{pmatrix} \,~0 & ~\mathbb{1}\,\\ \!-\mathbb{1} & ~0\,\end{pmatrix}\;,
\end{equation}
where $(\alpha_A,\beta^A)$ is a symplectic basis of $H^3(X,\mathbb Z)$, dual to a symplectic basis of three-cycles $(A^A,B_A)$ with
\begin{equation}
\int_{A^A}\alpha_B=\delta^A{}_B~,\qquad
\int_{B_A}\beta^B=-\delta_A{}^B ,
\end{equation}
and $\Pi$ is the period vector with entries
\begin{equation}
\cZ^A=\int_{A^A}\Omega~,\qquad
\cF_A=\int_{B_A}\Omega~.
\end{equation}
As already mentioned above, the index $A$ has a range $A=0,\ldots, h^{2,1}(X)$ since the periods provide homogeneous coordinates on the complex structure moduli space, reflecting the fact that $\Omega$ is defined only up to an overall complex rescaling. \\[2mm]
{\bfseries The flux superpotential.} From the Gukov--Vafa--Witten formula~\cite{Gukov:1999ya}, the flux superpotential, to leading order in $\alpha'$, has the form
\begin{equation}\label{Wflux}
W = w\int_X H\wedge \Omega
= w\left(n_A \cZ^A - m^A \cF_A\right)
= w\, \mu^T \eta \,\Pi \;,\qquad
\mu=\begin{pmatrix} n_A \\ m^A \end{pmatrix},
\end{equation}
with NSNS three-form flux $H\in H^3(X,\mathbb Z)$ given by
\begin{equation}
H = n_A\,\alpha_A - m^A\,\beta^A \, .
\end{equation}
Here $n_A,m^A\in\mathbb{Z}$ are the flux integers, defined (up to the usual $2\pi$ and $\alpha'$ convention factors which are absorbed into the prefactor $w$) by the periods of $H$ over the symplectic basis of three-cycles $(A^A,B_A)$, that is,
\begin{equation}
n_A = \int_{A^A} H \,, \qquad
m^A = \int_{B_A} H \, .
\end{equation}
The prefactor $w$ is of order one in string units and can be obtained from dimensional reduction; its precise value will not be important for the present discussion. \\[2mm]
{\bfseries The moduli K\"ahler potential.} To leading order, the moduli K\"ahler potential has the structure,
\begin{equation} \label{eq:kahler_tree}
    \cK = K^{(s)} + K^{(t)} + K\;,
\end{equation}
where $K$ is the complex structure K\"ahler potential and the first two terms are explicitly given by
\begin{eqnarray}
 K^{(s)}&=&-\log(S+\bar{S})\\
 K^{(t)}&=&-\log\left(\frac{1}{12}d_{ijk}(T^i+\bar{T}^i)(T^j+\bar{T}^j)(T^k+\bar{T}^k)\right)=-\log(\cV)\; , 
\end{eqnarray} 
with the CY volume $\cV=\frac{1}{6}d_{ijk}t^it^jt^k$ and the CY triple intersection numbers $d_{ijk}$. For our discussion, the complex structure K\"ahler potential $K$ is crucial and this can be written in terms of the period vector as
\begin{equation}\label{Kcs}
    K=-\log\left(i\int_X\Omega\wedge\bar{\Omega}\right)=-\log\left[i\left(\bar{\cZ}^A\cF_A-\cZ^A\bar{\cF}_A\right)\right]=
 -\log\left(-i\Pi^\dagger\eta\Pi\right)\;.
\end{equation}
The flux superpotential $W$ in Eq.~\eqref{Wflux} and the complex structure K\"ahler potential $K$ have been written in terms of the homogeneous coordinates $\cZ^A$, reflecting the projective nature of the complex structure moduli space. However, physical quantities must ultimately be expressed in terms of local coordinates on moduli space. To this end, we define the affine superpotential $\hat W(Z)$ and K\"ahler potential $\hat K(Z,\bar Z)$, which are related to their homogeneous counterparts by a K\"ahler transformation,
\begin{equation}\label{What}
     \hat{W}=\frac{W}{\cZ^0}\;,\qquad
     \hat{K}=K+\log(|\cZ^0|^2)\; .
\end{equation}
This transformation leaves the scalar potential and all physical observables invariant, while allowing us to perform explicit computations in terms of the physical moduli $Z^a$.\\[2mm] 
{\bfseries Flux vacua.} In this paper, we study solutions to the complex structure F-term equations
\begin{equation}\label{Ftermeqs}
F_a \equiv \hat W_a+\hat K_a\hat W \stackrel{!}{=}0 \, ,
\end{equation}
where $\hat W$ is the flux superpotential, $\hat W_a=\partial\hat W/\partial Z^a$ and $\hat K_a=\partial\hat K/\partial Z^a$. At a solution of these equations, and in the absence of any superpotential contributions other than the flux term, the scalar potential becomes
\begin{equation}\label{V_0}
V\big|_{F_a=0} = e^{\cK}\!\left(\cK^{S\bar S}\cK_S\cK_{\bar S} + \cK^{i\bar j}\cK_i\cK_{\bar j} -3 \right)|W|^2 .
\end{equation}
Using the tree-level K\"ahler potential~\eqref{eq:kahler_tree}, one finds
$\cK^{S\bar S}\cK_S\cK_{\bar S}=1$ and $\cK^{i\bar j}\cK_i\cK_{\bar j}=3$. As a result, the bracket in Eq.~\eqref{V_0} does not vanish, the no-scale mechanism is not realised and we have a residual scalar potential,
\begin{equation}\label{V0}
  V\big|_{F_a=0} =\frac{|U_0|^2}{2s\mathcal{V}}\;,
\end{equation}
where $U_0$ has been defined in Eq.~\eqref{U0}.
On its own, this potential induces a runaway behaviour towards $s\to\infty$ and $\cV\to\infty$, rather than a stabilised vacuum. 
One possible way to avoid this would be to find solutions to $F_a=0$ with $U_0=W_0=0$, so that the scalar potential~\eqref{V0} vanishes. However, our first no-go theorem below shows that such solutions do not exist in regions of moduli space where the metric derived from the K\"ahler potential~\eqref{Kcs} is non-singular.\\[2mm]
{\bf Heterotic moduli stabilisation.} Suppose we would like to stabilise all moduli by a combination of flux and non-perturbative effects~\cite{deCarlos:2005kh,Cicoli:2013rwa}, using a superpotential of the form~\eqref{Wfull} with the non-perturbative part schematically given by
\begin{equation}\label{Wnp}
  W_{\rm np}=w_{\rm np}\left(Ae^{-aT}+Be^{-bS}+\cdots \right) \; .
\end{equation}
Here $w_{\rm np}$ is a dimensionful quantity expected to be of order one in string units. The term proportional to $e^{-bS}$ can arise from gaugino condensation~\cite{Derendinger:1985kk,Nilles:1990jv}, while $e^{-aT}$ corresponds to a worldsheet instanton contribution~\cite{Witten:1996bn,Witten:1999eg}. Working in the large radius limit, $t={\rm Re}(T)\gg 1$, ensures that $|e^{-aT}|$ is exponentially suppressed. The overall magnitude of the term $|A e^{-aT}|$, however, depends on the Pfaffian prefactor $A$. While its dependence on the complex structure and bundle moduli can be computed in explicit examples~\cite{Buchbinder:2002ic,Buchbinder:2002pr,Buchbinder:2016rmw,Buchbinder:2019hyb,Braun:2007xh,Braun:2007tp,Buchbinder:2017azb}, there are currently no methods, short of a full numerical calculation, to determine its overall normalisation.\\[2mm]
However, it is reasonable to assume that higher order instanton terms, represented by the dots in Eq.~\eqref{Wnp}, are of order $A^n e^{-naT}$ for $n=2,3,\ldots$. Accuracy of the leading order instanton approximation then implies that $|A e^{-aT}|\ll 1$ and, hence, we should have $|W_{\rm np}|\ll 1$. For such small values of the non-perturbative superpotential to compete with the flux superpotential the quantity
\begin{equation}
    \frac{|U_{0,{\rm np}}|}{|U_0|}=\frac{|W_{0,{\rm np}}|}{|W_0|}\; ,
\end{equation}
where $U_{0,{\rm np}}=\langle e^{\hat{K}/2}\hat{W}_{\rm np}\rangle$ and $W_{0,{\rm np}}=\langle \hat{W}_{\rm np}\rangle$, evaluated at the solution to the complex structure F-term equations, should not be too small. More concretely, a constructive interplay between flux and non-perturbative superpotential which leads to a consistent stabilisation of all moduli is difficult to achieve if $|W_0|\gg 1$. The case $|W_0|={\cal O}(1)$ appears marginal: whether such values can facilitate consistent moduli stabilisation may depend on precise numerical coefficients in the instanton sector. Achieving values of $|W_0|$ that are moderately below unity improves the prospects for stabilisation, although parametrically small values, $|W_0|\ll 1$, would provide stronger control. At any rate, it is clear that the value of $|W_0|$ is crucial for heterotic moduli stabilisation. The main purpose of this paper is to investigate which values of $|W_0|$ can be realised in heterotic CY compactifications with flux and, in particular, whether values below unity can be achieved.\\[2mm]
It should be noted that $W_0$ and $W_{0,{\rm np}}$ are not individually K\"ahler invariant, but transform in the same way under K\"ahler transformations, so their relative size is meaningful once a frame is fixed. In the absence of a reliable determination of the non-perturbative superpotential, we evaluate $W_0$ in a natural frame defined by the period vector, following Ref.~\cite{Candelas:1990pi}, which is tied to the integral symplectic basis and flux quantisation. In this frame, comparing $|W_0|$ to unity provides a useful measure of whether the flux contribution can be sufficiently suppressed.\\[2mm]
{\bf Fluxes in Type IIB string theory.} For orientation, it is useful to recall the analogous situation in type~IIB string theory. In this case, the flux superpotential at leading order is given by 
\begin{equation} \label{eq:gvw_formula_iib}
    W=w \int_{X} (H+iSF) \wedge \Omega = w\,(\mu^T+i\bar{S}\rho^T)\,\eta\,\Pi\;, \qquad 
    \rho=\begin{pmatrix} a_A \\ b^A \end{pmatrix},
\end{equation}
where $\mu$ denotes the flux vector associated with the NSNS three-form $H$, as in Eq.~\eqref{Wflux}, while $\rho$ encodes the flux integers of the RR three-form $F$. Evidently, the type~IIB flux superpotential depends on twice as many flux integers as its heterotic counterpart, leading to a substantially larger discretuum of flux choices. Moreover, the explicit dependence on the axio-dilaton implies that one must now solve the coupled F-term equations
\begin{equation}
    F_a:=\hat{W}_a+\hat{K}_a\hat{W}\stackrel{!}{=}0\;,\qquad
    F_S:=\hat{W}_S+\hat{K}_S\hat{W}\stackrel{!}{=}0\; .
\end{equation}
The remaining fields, namely the K\"ahler moduli $T^i$, do not appear in the superpotential and satisfy the identity
$\cK^{i\bar j}\cK_i\bar{\cK}_{\bar j}=3$.
As a result, the no-scale cancellation is realised and the scalar potential vanishes at the flux vacuum, $ \langle V\rangle_{F_a=F_S=0}=0$, in contrast to the heterotic case, cf.\ Eq.~\eqref{V_0}.
In short, our conclusions concerning heterotic flux potentials cannot be directly transferred to generic type~IIB compactifications. However, if the RR flux is switched off, the type~IIB superpotential~\eqref{eq:gvw_formula_iib} has the same form as the heterotic superpotential~\eqref{Wflux}. In this special case, our conclusions do apply to type~IIB compactifications as well.

\section{No-go theorems} \label{sec:no-gos}
We have seen in the previous section that solving the complex structure F-term conditions in the presence of a heterotic flux superpotential $W$ leads to a residual potential term~\eqref{V_0} at leading order which can cause runaway behaviour for the dilaton and the volume modulus. As we have mentioned, one way to avoid this is to find solutions to the F-term equations~\eqref{Ftermeqs} with $W_0=\langle\hat{W}\rangle_{F_a=0}=0$ or, equivalently, solutions to $d\hat{W}=\hat{W}=0$.\\[2mm]
{\bfseries First no-go theorem.} Our first no-go theorem states that it is impossible to find viable solutions of this kind.
\begin{thm}[No-go theorem 1] \label{thm:nogo-1}
At solutions to $dW=0$ (which is equivalent to $d\hat{W}=\hat{W}=0$), where $W$ is the (non-zero) heterotic flux superpotential from Eq.~\eqref{Wflux} (and $\hat{W}$ is its affine counterpart from Eq.~\eqref{What}), the moduli space metric computed from the K\"ahler potential~\eqref{Kcs} is singular. 
\begin{proof}
The assumption $dW=0$ means that  $\frac{\partial W}{\partial \cZ^A}=0$ for all $A=0,\ldots ,h^{2,1}(X)$. It is not hard to show that this is equivalent to
\begin{equation}\label{Fmn}
\cF_{AB}m^B=n_A\qquad\mbox{where}\qquad\cF_{AB}=\frac{\partial^2\cF}{\partial\cZ^A\partial\cZ^B}
\end{equation}
is the period matrix. Since the fluxes $m^B$ and $n_A$ are real this implies ${\rm Im}(\cF_{AB})m^B=0$, so that the imaginary part of the period matrix is singular. (This conclusion can be avoided if all $m^B=0$ but then Eq.~\eqref{Fmn} implies that all $n_A=0$ as well. This means the flux superpotential vanishes, a case which is not of interest.) The well-known identity~\cite{Candelas:1990pi}, 
\[
K_{AB}:=\frac{\partial^2 K}{\partial\cZ^A\partial\cZ^B}=-2\,\text{Im}(\cF_{AB})
\]
then shows that the homogeneous version of the K\"ahler metric, $K_{AB}$, is also singular. It is not hard to prove that this implies the actual moduli space K\"ahler metric
\[
 \hat{K}_{a\bar{b}}=\frac{\partial^2 \hat{K}}{\partial Z^2\partial\bar{Z}^{\bar{b}}}\; .
\]
is singular as well, as claimed.
\end{proof}
\end{thm}\noindent
In conclusion, we cannot set the residual flux potential~\eqref{V_0} to zero. However, we can still hope that $|U_0|$ is small enough for specific flux choices so that non-perturbative effects can compete with~it. This requires solutions to the F-term equations~\eqref{Ftermeqs} with $|W_0|$ sufficiently small. Calculating $|W_0|$ requires knowledge of the period vector $\Pi$ or the pre-potential $\cF$ which has to be determined individually for each CY manifold. However, general statements for all CY manifolds may be possible in some limits in complex structure moduli space where generic expressions for the pre-potential can be written down.\\[2mm]
{\bfseries The large complex structure limit.} One such limit of interest is the large complex structure limit in which the pre-potential, up to exponential terms which are neglected, can be written as,
\begin{equation} \label{Flcs}
    \cF=-\frac{1}{6}\frac{\tilde{d}_{abc}\cZ^a\cZ^b\cZ^c}{\cZ^0}
    + \frac{1}{2}\tilde{a}_{ab} \cZ^a \cZ^b 
    + \tilde{b}_a \cZ^a\cZ^0 + \frac{1}{2} \tilde{c} (\cZ^0)^2\; .
\end{equation}
Here, $\tilde{d}_{abc}$ are the triple intersection numbers of the mirror CY $\tilde{X}$ of $X$
and the coefficients $\tilde{a}_{ab}$, $\tilde{b}_a$ and $\tilde{c}$ are given by
\begin{equation} \label{:b_c_coeff}
     \tilde{a}_{ab} = \begin{cases}
        \tilde{d}_{aab} &\text{for}\quad a\geq b\\
        \tilde{d}_{abb} &\text{for}\quad a < b
    \end{cases}\;,\qquad 
    \tilde{b}_a = \frac{1}{24} c_{2,a}(\tilde{X})\;,\qquad \tilde{c} = \frac{-i\,\zeta(3)}{(2\pi)^3} \eta(\tilde{X})\; ,
\end{equation}
where $c_2(\tilde{X})$ and  $\eta(\tilde{X})$ are the second Chern class and Euler number of $\tilde{X}$, respectively. Inserting the large complex structure pre-potential~\eqref{Flcs} into the general form of the flux superpotential~\eqref{Wflux} (and writing the result in affine coordinates) gives
\begin{equation}\label{Wlcs}
\hat{W}=w\left[n_0' - m^0c +i\,n_a'Z^a-\frac{1}{2}\tilde{d}_{abc}m^aZ^bZ^c+\frac{i\,m^0}{6}\tilde{d}_{abc}Z^aZ^bZ^c\right]\; ,
\end{equation}
where
\begin{equation} \label{nmeqs}
    n_0' = n_0  - m^a\tilde{b}_a\;,\qquad
    n_a' = n_a - m^0\tilde{b}_a - m^b\tilde{a}_{ba}\;.
\end{equation}
Here, we have performed a field re-definition $Z^a\mapsto iZ^a$, so that, writing $Z^a=z^a+i\chi^a$, the real part $z^a$ corresponds to the geometrical modulus and $\chi^a$ to the axion. The complex structure K\"ahler potential in the large complex structure limit is obtained by inserting the pre-potential~\eqref{Flcs} into Eq.~\eqref{Kcs}. Using the same conventions for the fields as above this leads to
\begin{equation}\label{Klcs}
 \hat{K}=-\log\left(\frac{4}{3}\tilde{\kappa}+k\right)\;,\qquad \tilde{\kappa}=\tilde{d}_{abc}z^az^bz^c\; ,
 \qquad k=i(\tilde{c}-\tilde{c}^*)\; .
\end{equation}
For this K\"ahler potential to be well-defined we must have $\tkap>-3k/4$.\\[2mm]
{\bfseries Second no-go theorem.} The above expressions for $\hat{W}$ and $\hat{K}$ are sufficiently concrete to attempt to solve the F-term equations~\eqref{Ftermeqs} analytically and compute the resulting values of $|W_0|$.
Unfortunately, this leads to the following no-go theorem.
\begin{thm}[No-go theorem 2] \label{thm:nogo-2} Solutions to the F-term equations~\eqref{Ftermeqs} for a (non-zero) large complex structure superpotential~\eqref{Wlcs} and large complex structure K\"ahler potential~\eqref{Klcs} satisfy
\begin{equation}\label{nogo2ineq0}
\left.\begin{array}{lcl} \left|{\rm Im}(W_0)\right|\\[2mm]\left|{\rm Re}(W_0)\right|
\end{array}\right\}
=\frac{2|w|}{3}\left(1+\frac{3k}{4\tkap}\right)\left\{\begin{array}{ccl}|m^0|\tkap&\mbox{for}&m^0\neq 0\\[2mm]|\tilde{\kappa}(m)|^{1/3}\tkap^{2/3}&\mbox{for}&m^0=0\end{array}\right.\;,
\end{equation}
and
\begin{equation}\label{nogo2ineq}
\left.\begin{array}{lcl} \left|{\rm Im}(U_0)\right|\\[2mm]\left|{\rm Re}(U_0)\right|
\end{array}\right\}
=|w|\sqrt{\frac{1}{3}\left(1+\frac{3k}{4\tkap}\right)}\left\{\begin{array}{ccl}|m^0|\tkap^{1/2}&\mbox{for}&m^0\neq 0\\[2mm]|\tilde{\kappa}(m)|^{1/3}\tkap^{1/6}&\mbox{for}&m^0=0\end{array}\right.\;,
\end{equation}
Here, $\tilde{\kappa}$ is the expression defined in Eq.~\eqref{Klcs} evaluated for the solution to the F-term equations and $\tilde{\kappa}(m):=\tilde{d}_{abc}m^am^bm^c$ is a non-zero integer. 
\begin{proof}
    The proof involves solving the F-term equations~\eqref{Ftermeqs} with the large complex structure expressions~\eqref{Wlcs} and \eqref{Klcs} inserted, using very special geometry identities and splitting the fields as $Z^a=z^a+i\chi^a$ into real and imaginary parts. A case distinction between $m^0=0$ and $m^0\neq 0$ is required. For the $m^0 = 0$ case, while $\tilde{\kappa}(m)$ may well become zero for certain flux choices, the detailed proof shows this implies the vanishing of all fluxes. Hence, this case is excluded by our assumption of a non-zero flux superpotential. The details of the proof are a bit technical and a full account is contained in Appendix~\ref{app:nogo2}.
\end{proof}
\end{thm}\noindent
How should the above theorem be interpreted physically? The large complex structure limit implies that the corresponding mirror volume is large, that is, $\tkap = \tilde{d}_{abc}z^az^bz^c \gg 1$. The loop corrections to the leading-order cubic term in Eq.~\eqref{Flcs} then become small in this limit, and in particular we have $|k|/\tkap \ll 1$. In this case, $\left(1+\frac{3k}{4\tkap}\right) \sim 1$ and Eq.~\eqref{nogo2ineq0} implies that $|W_0|\gg 1$. This means that non-perturbative effects cannot compete with the flux superpotential, and that it is unlikely that consistent heterotic moduli stabilisation can be achieved in the large complex structure limit. Frequently, a large complex structure pre-potential of the form
\begin{equation} \label{Flcs0}
 \cF=-\frac{1}{6}\frac{\tilde{d}_{abc}\cZ^a\cZ^b\cZ^c}{\cZ^0}
\end{equation}
is used, effectively neglecting the sub-leading terms in Eq.~\eqref{Flcs} (or, formally, setting $\tilde{a}_{ab}=\tilde{b}_a=\tilde{c}=0$ in Eq.~\eqref{Flcs}). Theorem~\ref{thm:nogo-2} can be specialised to this case by setting $k=0$ and this shows that the above conclusions remain valid for a pre-potential of the form~\eqref{Flcs0}.\\[2mm]
We should, therefore, study the problem away from the large complex limit, and consider other regions in the complex structure moduli space to investigate the possibility of small $|W_0|$. In the remainder of the paper, we will do this for a number of explicit examples.

\section{Heterotic flux in one-parameter models} \label{sec:one-param}
The no-go theorem~\ref{thm:nogo-2} shows that we should investigate local supersymmetric vacua away from the large complex structure limit, for a chance of finding vacua with small $|W_0|$. Since the period vector $\Pi$ consists of complicated functions which need to be computed for each CY manifold this is difficult to analyse generically. Instead, we will consider examples (or families of examples) of CY manifolds where the period vector has been determined analytically. In this section, we start with examples of one-parameter models, that is, CY manifolds $X$ with $h^{2,1}(X)=1$. The primary example is the mirror quintic introduced in the seminal work~\cite{Candelas:1990rm} as well as mirror duals of other Fermat hypersurfaces~\cite{Strominger:1985it,Klemm:1992tx} where the complex structure moduli dependence of the prepotential is known. The period functions are typically too complicated for exact analytic calculations but, as we will see, progress can be made by numerical methods and suitable approximations.

\subsection{Manifolds and periods} \label{subsec:mfds}
{\bfseries Definition of manifolds.} Our basic examples are the mirror duals $X$ of CY hyper-surfaces $\tilde{X}$, described as the vanishing locus in the weighted projected space $\mathbb{P}_{\nu_0,\ldots ,\nu_4}$ of a quasi-homogeneous polynomial of the Fermat type~\cite{Strominger:1985it,Klemm:1992tx},
\begin{equation}\label{p}
    p = \sum_{i=0}^4 x_i^{n_i} -k\psi \prod_{i=0}^4x_i\stackrel{!}{=} 0\;,
\end{equation}
where $k$ is an integer, the powers $n_i$ are given by $n_i=k/\nu_i$ and $\psi$ is a complex parameter. The vanishing of the first Chern class imposes restrictions on these integers and this leads to four cases, for $k=5,6,8,10$, with powers $n_i$ and weights $\nu_i$ as given in Table~\ref{tab:fermat_folds}. 

\begin{table}[!ht]
    \centering\small
    \begin{tabular}{|c|c|c|c|c|c|}
        \hline
        $k$ & $G$ & $\text{Ord}(G)$ & $n_i$ & $\nu_i = k/n_i$ & $\gamma = k\prod_{i=0}^4(\nu_i^{-\nu_i/k})$\\
        \hline
        5 & $Z_5 \times Z_5 \times Z_5$ & $5^3$ & (5,5,5,5,5) & (1,1,1,1,1) & 5\\ 
        6 & $Z_3 \times Z_6 \times Z_6$ & $3 \times 6^2$ & (3,6,6,6,6) & (2,1,1,1,1) & $6\times2^{-1/3}$\\ 
        8 & $Z_8 \times Z_8 \times Z_2$ & $2\times 8^2$ & (2,8,8,8,8) & (4,1,1,1,1) & 4\\ 
        10 & $Z_{10} \times Z_{10}$ & $10^2$ & (2,5,10,10,10) & (5,2,1,1,1) & $10\times5^{-1/2}\times2^{-1/5}$\\
        \hline
    \end{tabular}
    \caption{\small A list of useful quantities for the Fermat-type CY three-fold families with $h^{1,1}(X) =1$. Here $k$ is the smallest common multiplet on $n_i$ where $n_i$ are the polynomial order of each coordinate, $G$ is the orbifold action, $\text{Ord}(G)$ is the order of the group $G$, $\nu_i$ are the weights of $\mathbb{P}^4$ coordinates and $\gamma = k\prod_{i=0}^4(\nu_i^{-\nu_i/k})$ is a useful quantity defined presently.}
    \label{tab:fermat_folds}
\end{table}

\noindent These four manifolds $\tilde{X}$ have a single K\"ahler modulus, so that $h^{1,1}(\tilde{X})=1$, as well as a discrete symmetry group $G$ which is listed in Table~\ref{tab:fermat_folds}. We are interested in the mirrors $X$ of these manifolds which can be constructed by forming the quotient of $\tilde{X}$ by $G$. These mirrors have a single complex structure parameter, that is $h^{2,1}(X)=1$, which corresponds to the parameter $\psi$ in Eq.~\eqref{p}.\\[2mm]
There are three points of interest in the one-dimensional complex-structure moduli spaces of the mirror Fermat hypersurfaces. Firstly, the point $\psi = 1$ corresponds to a singular manifold where the hypersurface fails to be transverse, known as the conifold point. The point $\psi\rightarrow\infty$ corresponds to the large complex structure limit. Finally, $\psi = 0$ is the mirror dual of the Fermat manifold defined as the zero locus of the polynomial~\eqref{p} with $\psi=0$.\\[2mm]
{\bfseries Fundamental period.} To describe the period vector we can start with the fundamental period  $\varpi_0$ in the region $|\psi|>1$ defined as
\begin{equation}
    \varpi_0(\psi) = \int_{\mathcal{B}} \Omega(\psi)\;, 
\end{equation}
where $\Omega(\psi)$ is the holomorphic $(3,0)$-form, represented in a specific way~\cite{Candelas:1990rm,Klemm:1992tx} and $\mathcal{B}$ is known as the fundamental three-cycle.\\[2mm]
The three points of interest described above imply that there are two natural regions, $|\psi|>1$ and $|\psi|<1$, in each complex structure moduli space.
Explicit expressions for $\varpi_0(\psi)$ in the region $|\psi|>1$ can be obtained as power series in $1/\psi$, which can be analytically continued to the region $|\psi|<1$.\\[2mm]
{\bfseries Period vector.} We can define further periods by
\begin{equation} \label{eq:comp_basis}
    \varpi_j(\psi) = \varpi_0(\alpha^j\psi)\;, \qquad \alpha = \exp\left(\frac{2\pi i}{k}\right)\;,
\end{equation}
where $j=0,\ldots ,k-1$ in the region $|\psi| <1$. There are $k-4$ linear relations between those periods and, hence, there are only four linear independent ones which can be taken to be $(\varpi_0,\varpi_1,\varpi_2,\varpi_{k-1})$. The period vector $\Pi$ can then be written as
\begin{equation}\label{pidef}
\Pi=\left(\begin{array}{l}\cF_0\\\cF_1\\\cZ^a\\\cZ^1\end{array}\right)=M\varpi\qquad\mbox{where}\qquad
\varpi=-\frac{(2\pi i)^3}{\text{ord}(G)}\left(\begin{array}{c}\varpi_{2}\\\varpi_1\\\varpi_0\\\varpi_{k-1}\end{array}\right)\; ,
\end{equation}
where $\text{ord}(G)$ indicates the order of the $G$ action in each of the families and $M$ is a constant $4\times 4$ `monodromy matrix'. These periods can be analytically-continued to the $|\psi| > 1$ region and detailed expressions for $\varpi_j$ in all regions as well as the monodromy matrices $M$ can be found in Appendix~\ref{sec:periods_full}. These expression are used to find the period vector $\Pi$ from Eq.~\eqref{pidef} which forms the basis of our subsequent numerical calculations.

\subsection{Systematic searches for flux vacua} \label{subsec:num_search}

Given the period vector $\Pi$ for our one-parameter examples, we compute the flux superpotential and K\"ahler potential by inserting $\Pi$ into Eqs.~\eqref{eq:gvw_formula_iib} and \eqref{Kcs}, respectively, thereby expressing these quantities as functions of the complex structure modulus $\psi$. We then solve the F-term equation
\begin{equation}\label{Fterms1}
    F_\psi = \hat{W}_\psi + \hat{K}_\psi \hat{W}\stackrel{!}{=}0 \;.
\end{equation}
For the numerical analysis, Eq.~\eqref{Fterms1} is expanded to order $15$ in $\psi$, for $|\psi|<1$, or in $1/\psi$, for $|\psi|>1$. The resulting truncated equation is separated into real and imaginary parts and solved using a variant of Newton's method. In practice, we employ \verb|gsl_multiroot_fsolver| from the \texttt{gsl} C library, which updates a solution of $f(x)=0$ according to
\begin{equation}\label{Newton}
    x \mapsto x - \epsilon J(x)^{-1} f(x)\;,\qquad
    J_{ij}(x) = \frac{\partial f_i}{\partial x_j}(x)\; ,
\end{equation}
with step size controlled by $\epsilon$.
For each flux choice, we initialise the algorithm with a set of random starting values for $\psi$, drawn uniformly from the relevant region. Solutions to the F-term equation~\eqref{Fterms1}, together with the corresponding values of $W_0$ and $U_0$, are collected. This procedure is repeated for all flux configurations within the range $|n_A|,|m^A|\leq 32$. For each model, we then identify the flux choice and solution $\psi_{\rm min}$ that yield the smallest value of $|W_0|$.\\[2mm]
Our numerical scans indicate that in the region $|\psi|>1$, solutions to the F-term equations exist but are associated with large values $|W_0|\gtrsim \mathcal{O}(10^2)$. In contrast, in the region $|\psi|<1$, we find solutions with $|W_0|\sim \mathcal{O}(1)$. The best results for the four CY manifolds under consideration are summarised in Table~\ref{tab:fermat_vacua_new}, while the distribution of $U_0$ and $W_0$ values obtained in the scan is shown in Figures~\ref{fig:c} and ~\ref{fig:psi_plots}, respectively.

\begin{table}[!h]
\centering\small
\begin{tabular}{|c|c|c|c|c|c|c|}
    \hline
    $k$ & $(n_0,n_1,m^0,m^1)$ & $\psi_{\min}$ & $W_0$ & $|W_0|$ & $U_0$ & $|U_0|$ \\
    \hline
    \hline
    5 & $(-3,0,-5,1)$ & $0.465 + 0.338i$ & $-0.283 - 0.111i$ & $0.304$ & $-1.34 - 0.527i$ & $1.44$\\
    \hline
    6 & $(0,-1,5,1)$ & $0.565 + 0.542i$ & $-0.662 - 0.500i$ & $0.829$ & $-1.48 + 1.12i$ & $1.85$\\
    \hline
    8 & $(0,-1,5,1)$ & $0.777+0.515i$ & $-0.848 + 0.808i$ & $1.17$ & $-1.49 + 1.42i$ & $2.06$\\
    \hline
    10 & $(0,0,-3,-1)$ & $0.438 + 0.142i$ & $-1.23 + 4.16i$ & $4.34$ & $-0.755-2.50i$ & $2.61$\\
    \hline
\end{tabular}
\caption{Heterotic flux superpotential results for the mirror Fermat hypersurfaces in Table~\ref{tab:fermat_folds}, using period expressions valid in the region $|\psi|<1$. For each model, we perform a scan over flux integers in the range $|n_A|,|m^A|\leq 32$ and display the flux choice $(n_0,n_1,m^0,m^1)$ that admits a supersymmetric minimum at $\psi_{mr min}$ with the smallest value of $|W_0|$. The corresponding values of $\psi_{\rm min}$, $W_0$, $|W_0|$, $U_0$, and $|U_0|$ are also shown.}
\label{tab:fermat_vacua_new}
\end{table}
\begin{figure}[h!]
    \centering
    \begin{subfigure}[t]{0.45\linewidth}
        \centering
        \includegraphics[scale=0.65]{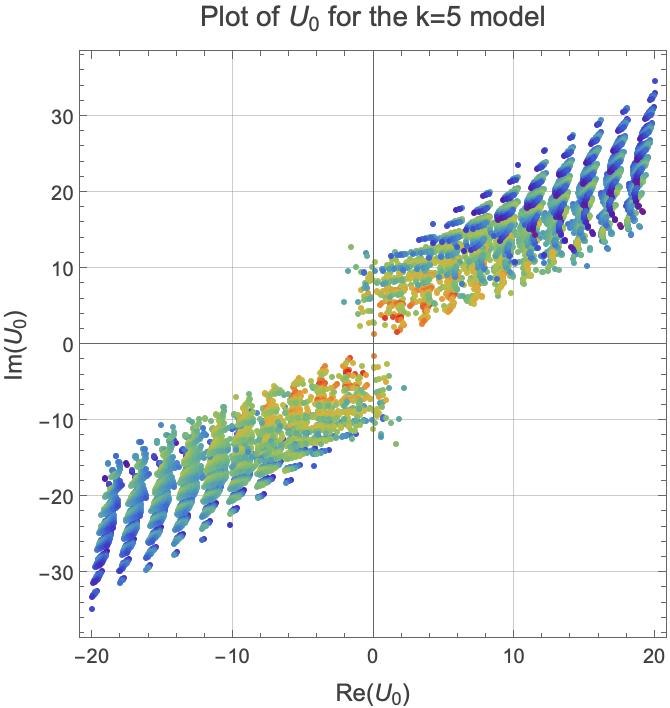}
    \end{subfigure}
   \begin{subfigure}[t]{0.45\linewidth}
        \centering
        \includegraphics[scale=0.62]{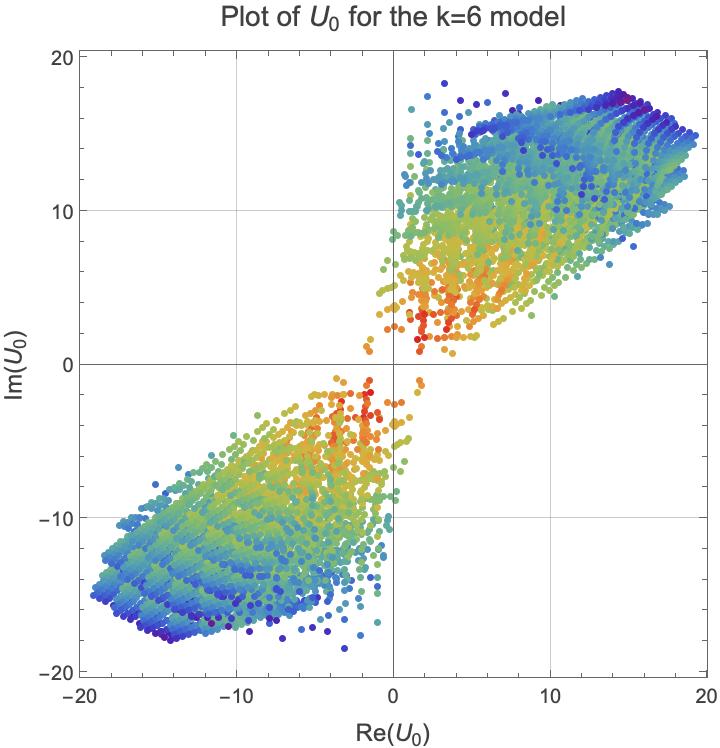}
    \end{subfigure}
    \vskip 4mm
    \begin{subfigure}[t]{0.45\linewidth}
        \centering
        \includegraphics[scale=0.65]{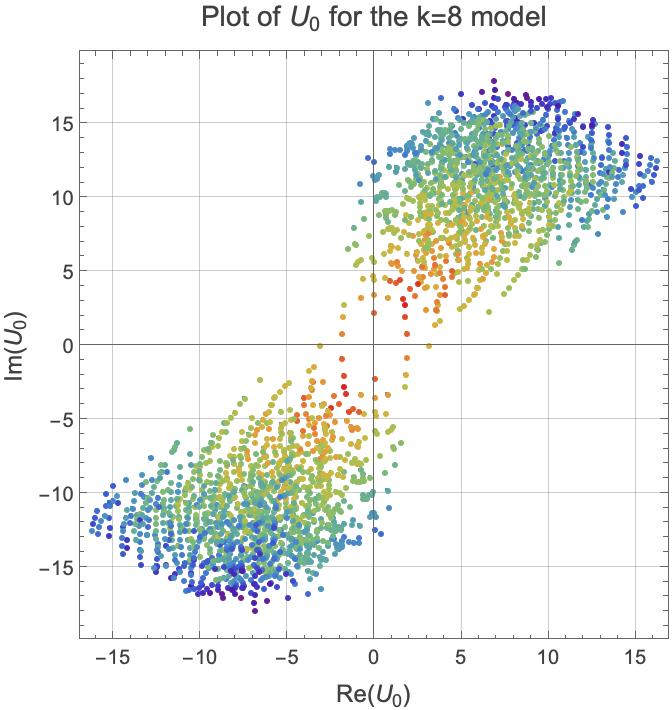}
    \end{subfigure}
   \begin{subfigure}[t]{0.45\linewidth}
        \centering
        \includegraphics[scale=0.65]{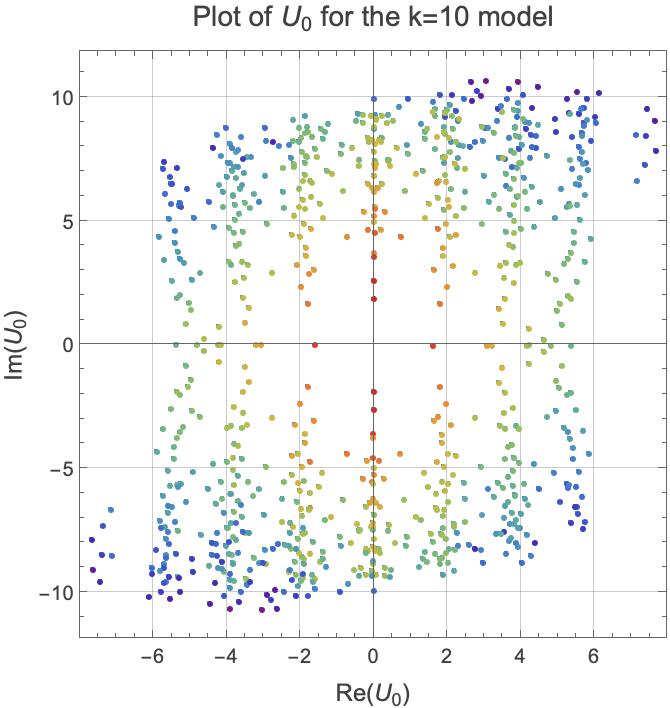}
    \end{subfigure}
    \vskip 3mm
    \caption{Plots of the values of $U_0$, defined in Eq.~\eqref{U0}, obtained from a systematic search for flux integers in the range $|n_A|,|m^A| \leq 10$. The four panels correspond to the four CY manifolds in Table~\ref{tab:fermat_folds}. The colours of the points represent the size, $\sum_A(|n_A|+|m^A|)$, of the flux vector, with red (blue) corresponding to the smallest (largest) values. The distributions of the values of $U_0$ are qualitatively different in each of the four manifolds owing to the different period structures.}
    \label{fig:c}
\end{figure}
\begin{figure}[h!]
    \centering
    \begin{subfigure}[t]{0.45\linewidth}
        \centering
        \includegraphics[scale=0.65]{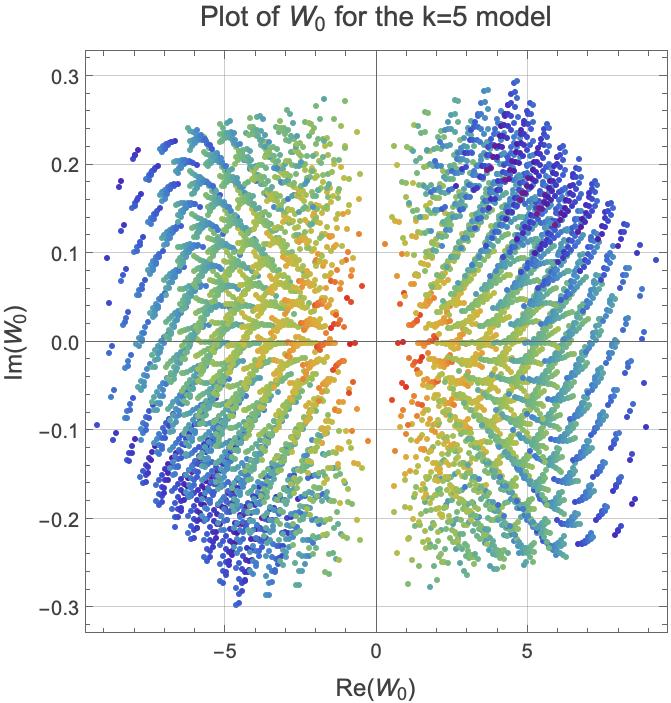}
    \end{subfigure}
   \begin{subfigure}[t]{0.45\linewidth}
        \centering
        \includegraphics[scale=0.65]{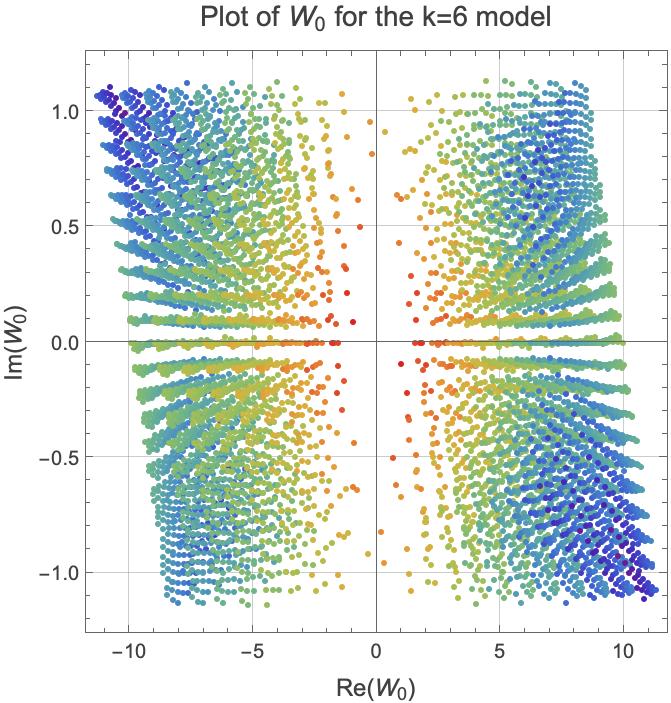}
    \end{subfigure}
    \vskip 4mm
    \begin{subfigure}[t]{0.45\linewidth}
        \centering
        \includegraphics[scale=0.625]{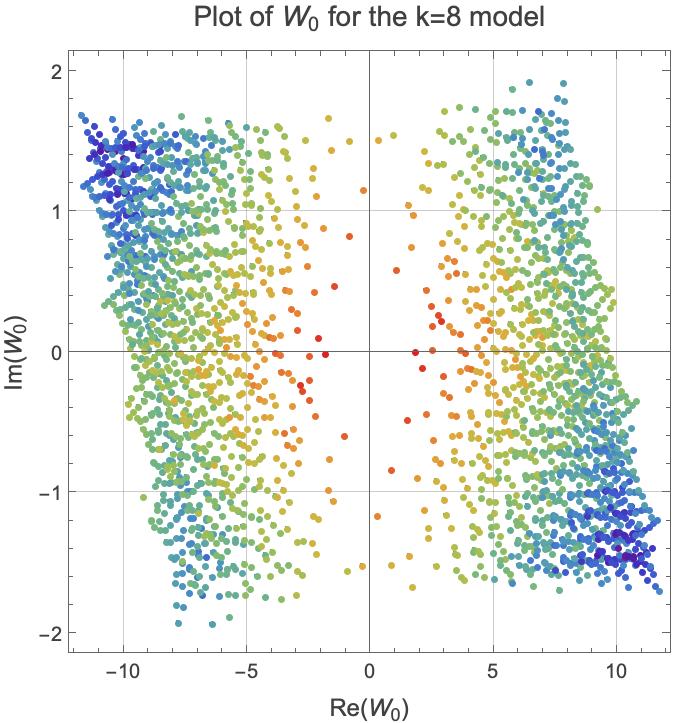}
    \end{subfigure}
   \begin{subfigure}[t]{0.45\linewidth}
        \centering
        \includegraphics[scale=0.65]{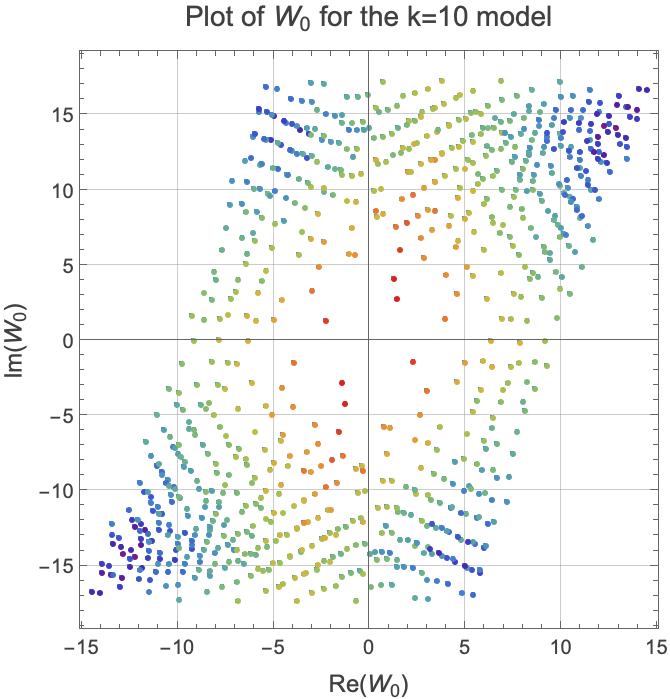}
    \end{subfigure}
    \vskip 3mm
    \caption{Plots of the values of $W_0$ from Eq.~\eqref{U0} obtained from a systematic search for flux integers in the range $|n_A|,|m^A| \leq 10$. The four panels correspond to the four CY manifolds in Table~\ref{tab:fermat_folds}. The colours of the points represent the size, $\sum_A(|n_A|+|m^A|)$, of the flux vector, with red (blue) corresponding to the smallest (largest) values.}
    \label{fig:psi_plots}
\end{figure}

\noindent From Table~\ref{tab:fermat_vacua_new}, the minimal values of $|W_0|$ within the search region $|n_A|,|m^A|\leq 32$ are of order one. As discussed earlier, such values may be marginally compatible with stabilisation of all moduli. Extending the scan to a larger region in flux space becomes computationally expensive. Instead, we explore a wider range of fluxes using a heuristic search based on genetic algorithms. 

\subsection{Heuristic searches for flux vacua}
Heuristic search methods, in particular genetic algorithms and reinforcement learning, have been successfully employed to explore large regions of the string landscape that are inaccessible to systematic scans~\cite{Harvey:2021oue,Krippendorf:2025mhp,Walden:2025cpf,Constantin:2021for,Abel:2021rrj,Abel:2023zwg}. In this work, we use genetic algorithms (GAs) to probe a wider range of flux integers beyond the limits of our systematic search. \\[2mm] 
To this end, we consider flux integers in the range $n_0,n_1,m^0,m^1 \in \{-2^{N-1}+1,\dots,2^{N-1}\}$, so that each integer can be represented by a binary string of length $N$. A flux configuration is then encoded as a bit string of total length $4N$, obtained by concatenating the four individual strings. The fitness function $f$ measures whether a solution to the F-term equation can be found and, if so, how small the corresponding $|U_0|$ value is. More precisely, if solutions to the F-term equation can be found for a given flux choice $(n_0,n_1,m^0,m^1)$, the fitness function is given by
\begin{equation}
 f(n_0,n_1,m^0,m^1)=-\min_{\psi:\,F_\psi=0}\left(\frac{|U_0|(\psi)}{U_{\rm max}}\right)\; ,
\end{equation}
where $U_{\rm max}=10$ is a reference scale used to normalise the fitness function and keep its values within a convenient numerical range. Otherwise, if no solution to the F-term equation can be found for a given flux choice $(n_0,n_1,m^0,m^1)$, we define
\begin{equation}
 g(n_0,n_1,m^0,m^1)=\min_{\psi\in \mathcal{S}(n_0,n_1,m^0,m^1)} |F_\psi|^2
\end{equation}
where $\mathcal{S}(n_0,n_1,m^0,m^1)$ denotes the set of candidate solutions obtained from the numerical root-finding procedure starting from different initial values of $\psi$.
We then define the fitness function that measures the degree to which the F-term equations fail to be satisfied
\begin{equation}
 f(n_0,n_1,m^0,m^1) = -\,\mathrm{clip}\big(g(n_0,n_1,m^0,m^1),\,g_{\rm min},\,g_{\rm max}\big)\;,
\end{equation}
where $\mathrm{clip}(x,a,b)$ denotes the operation that truncates $x$ to lie within the interval $[a,b]$, that is, values smaller than $a$ are set to $a$, values larger than $b$ are set to $b$ and values within the interval are kept unchanged. 
The cut-off values $g_{\rm min}$ and $g_{\rm max}$ are introduced to improve numerical stability and are chosen as $g_{\rm min}=10$ and $g_{\rm max}=10^4$.\\[2mm]
A flux configuration is considered \emph{viable} if $f(n_0,n_1,m^0,m^1)>-1$, corresponding to cases where solutions to the F-term equation exist with $|U_0|<10$.
\\[2mm]
{\bfseries GA search results.} We have carried out a GA search for $N=8$, that is, for flux integers in the range $n_0,n_1,m^0,m^1 \in \{-127,\dots,128\}$. GA runs are repeated until $\sim 75\%$ of the solutions identified in the systematic search (over the restricted range) are found. No smaller values of $|U_0|$ have been found in the larger search box, $n_0,n_1,m^0,m^1 \in \{-127,\dots,128\}$.
\section{Heterotic flux in a two-parameter model} \label{sec:two-param}
The minimal $|W_0|$ values we have been able to find for the one-parameter CY manifolds analysed in the previous section were of order one. As discussed such values may be marginally viable for heterotic moduli stabilisation but probably too large for comfort. With just four flux integers one-parameter models are somewhat restrictive and one can speculate that more complicated models with more flux integers might allow for smaller values of $|W_0|$. In order to test this we will now study a two-parameter model with six flux integers.

\subsection{A two-parameter model}
{\bfseries Definition of the manifold}. 
We start with a two-complex parameter family of hyper-surfaces defined as the zero locus of polynomials,
\begin{equation}
    p = x_1^8 + x_2^8 + x_3^4 + x_4^4 + x_5^4 - 8\psi\, x_1x_2x_3x_4x_5 - 2\phi\, x_1^4x_2^4\;,
\end{equation}
in the weighted projective space $\mathbb{P}^{(1,1,2,2,2)}$, where $\psi,\phi\in\mathbb{C}$ are the two complex structure parameters. After resolving  singularities this leads to a family of CY three-folds $\tilde{X}$ with symmetry $G \cong \mathbb{Z}_4 \times \mathbb{Z}_4 \times \mathbb{Z}_4$. The manifolds we are ultimately interested in are the mirrors $X$ of $\tilde{X}$
which are constructed as the quotients $X=\tilde{X}/G$.\\[2mm]
{\bfseries Moduli space.}
To analyse the moduli space, it is easier to enlarge $G$ to a symmetry group $\hat{G}$ which includes a $\mathbb{Z}_8$ symmetry generated by the element $g = (\alpha^{a_1},\alpha^{a_2},\alpha^{2a_3},\alpha^{2a_4},\alpha^{2a_5})$, which acts on the coordinates and the moduli space parameters as,
\begin{equation}
    (x_1,x_2,x_3,x_4,x_5;\psi,\phi) \mapsto (\alpha^{a_1}x_1, \alpha^{a_2}x_2, \alpha^{a_3}x_3, \alpha^{2a_4}x_4, \alpha^{2a_5}x_5; \alpha^{-a}\psi, \alpha^{-4a}\phi)\;,
\end{equation}
where $a = a_1 + a_2 + 2a_3 + 2a_4 + 2a_5$, $a_i\in \mathbb{Z}$ and $\alpha$ is an eighth root of unity. This additional symmetry acts non-trivially on $\psi$ and $\phi$ and it implies that the naïve moduli space must be  modded out by a $\mathbb{Z}_8$ symmetry generated by
\begin{equation}
    (\psi,\phi) \stackrel{g_0}{\mapsto} (\alpha\psi,-\phi)\;.
\end{equation}
This $\mathbb{Z}_8$-symmetry is important and facilitates generating the full set of periods functions.
In analogy with the mirror quintic, values of the moduli where the CY space becomes singular separate the moduli space into different regions. The details of these singular loci can be found in Ref.~ \cite{Candelas:1993dm}. They separate the moduli space into regions labelled by the signs of the expressions $|8\psi^4/(\phi\pm1)| - 1$ and $|\phi| - 1$.\\[2mm]
{\bfseries Fundamental period.} Similar to the mirror quintic case, to find the period vector we must first find the fundamental period $\varpi_0(\psi)$. It is defined as an integral of the holomorphic three-form $\Omega$ over the fundamental cycle $\cB$, that is,
\begin{equation}
    \varpi_0(\phi,\psi) = \int_{\cB}\Omega\;, \quad \left|\frac{8\psi^4}{\phi\pm1}\right|>1\;, \quad |\phi|>1\;.
\end{equation}
The evaluation of this integral can be carried out in the region $|8\psi^4/(\phi\pm1)|>1$ and $|\phi| > 1$. The result in the other regions of moduli space is obtained by analytic continuation and the details of this procedure can be found in  Appendix~\ref{subsec:period_full_two}.\\[2mm]
{\bfseries Period vector.} We can now define further periods by,
\begin{equation}
    \varpi_j(\psi,\phi) = \varpi_0(\alpha^j\psi,\alpha^{4j}\phi)\;, \quad \alpha = \exp\left(\frac{2\pi i}{8}\right)\;,
\end{equation}
where $j = 0,\dots,7$. There are two linear relations between these eight periods
and we can select six linearly independent ones. In terms of these, the period vector can be written as
\begin{equation} \label{eq:period_2_defn}
    \Pi(\psi) = \begin{pmatrix}
        \cF_0\\ \cF_1\\\cF_2\\ \cZ^0\\ \cZ^1\\\cZ^2
    \end{pmatrix}= M \varpi(\psi) \;,
    \quad \text{where}\quad
    \varpi = 
    \begin{pmatrix}
        \varpi_0\\ \varpi_1\\ \varpi_2\\ \varpi_3\\\varpi_4\\ \varpi_5
    \end{pmatrix}\;,
\end{equation}
and $M$ is a constant $6 \times 6$ `monodromy matrix'. Expressions for the periods $\varpi_j(\psi)$ and for $M$ can be found in Appendix~\ref{sec:periods_full}. As before, we use these expressions to calculate the period vector $\Pi$ in Eq.~\eqref{eq:period_2_defn} and subsequently analyse solutions to the F-term equations numerically.

\subsection{Searches for flux vacua}
Now we have two F-terms which, written in affine coordinates, are
\begin{equation}
    F_\psi = \hat{W}_\psi +\hat{K}_\psi\hat{W}\;,\quad
    F_\phi = \hat{W}_\phi +\hat{K}_\phi\hat{W}\;,
\end{equation}
and the subscripts $\phi$ and $\psi$ indicate partial differentiation.
As before, we compute the F-terms expanded to order 15 in $\psi$ and $\phi$ and solve the F-term equations numerically, using the method described around Eq.~\eqref{Newton}. This is done for all flux choices $(n_A,m^A)$ in a search box defined by $|n_A|,|m^A|\leq 8$ for the region $|8\psi^4/(\phi\pm1)| < 1$ and $|\phi| <1$ where period expansions have been computed in the literature~\cite{Berglund:1993ax}. The values of $W_0$ and $U_0$ obtained in this way have been plotted in Fig.~\ref{fig:regions_two_param_1}. The smallest value of $|U_0|$ we have been able to find within the search box is of order one and the details are provided in Table~\ref{tab:two_param}.

\begin{table}[!h]
\centering\small
\begin{tabular}{|c|c|c|c|c|c|}
    \hline
    $(n_0,n_1,n_2,m^0,m^1,m^2)$ & $(\psi_{\min},\phi_{\min})$ & $W_0$ & $|W_0|$ & $U_0$ & $|U_0|$\\
    \hline
    $(-2,-1,2,1,-2,0)$ & $(0.180{+}0.170i,\,-0.369{+}0.249i)$ & $5.84+2.63i$ & $6.40$ & $2.54+1.15i$ & $2.79$\\
    \hline
\end{tabular}
\caption{Solution to the F-term equations $F_\psi=F_\phi=0$ for the two-parameter model introduced in the text that yields the smallest value of $|W_0|$ within the flux range $n_0,n_1,n_2,m^0,m^1,m^2\in\{-8,\ldots,8\}$. Also shown are the corresponding values of $(\psi_{\min},\phi_{\min})$, $W_0$, $|W_0|$, $U_0$, and $|U_0|$.}
\label{tab:two_param}
\end{table}

\begin{figure}[H]
    \centering
    \begin{subfigure}[t]{0.45\linewidth}
        \centering
        \includegraphics[scale=0.65]{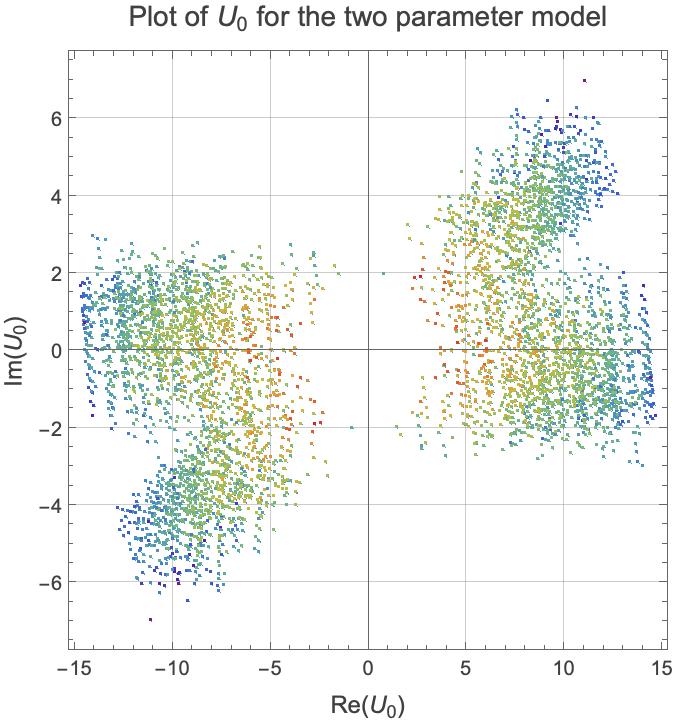}
    \end{subfigure}
   \begin{subfigure}[t]{0.45\linewidth}
        \centering
        \includegraphics[scale=0.65]{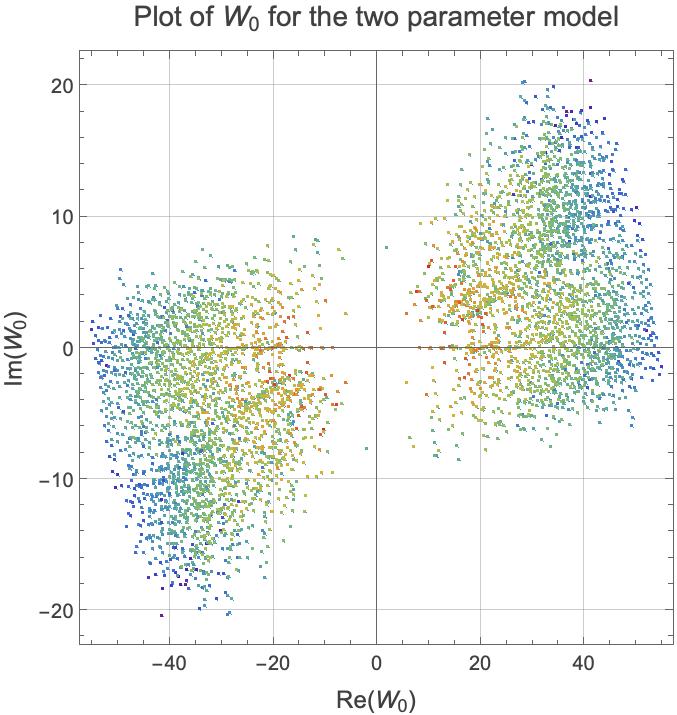}
    \end{subfigure}
    \caption{\small The plots show the values of $U_0$ and $W_0$ from Eq.~\eqref{V0} for the two-parameter model defined in the text, with fluxes in the range $|n_A|,|m^A| \leq 5$.} \label{fig:regions_two_param_1}
\end{figure}\noindent

\section{Conclusions} \label{sec:conclusion}
In this work we have explored moduli stabilisation using fluxes in heterotic string theories. Flux in heterotic theories is quite different from type IIB flux. Firstly, the heterotic flux superpotential does not depend on the axio-dilaton and this implies the absence of a no-scale structure. Also, additional effects must be considered to stabilise the axio-dilaton. Secondly, while type IIB flux originates from both the NSNS and RR three-forms, only the former is available in the heterotic case. As a result, the heterotic flux superpotential is much more restrictive, and only contains half the flux integers, compared to its IIB counterpart. It is well-known that small values $|W_0|$ of the flux superpotential at supersymmetric minima can be obtained in IIB string theory. However, due to the aforementioned differences, it is not clear which $|W_0|$ values can arise in the heterotic case. Small $|W_0|$ values are essential for heterotic moduli stabilisation and the main purpose of this paper has been to analyse if such values can be achieved.\\[2mm]
To guide our investigation we have first proved two no-go theorems. The first one asserts that supersymmetric Minkowski vacua of the flux superpotential do not exist in the part of the moduli space where the moduli space metric is non-singular. Essentially, this means that the residual heterotic flux potential~\eqref{V0}, cannot be set to zero. The second no-go theorem states that in the large complex structure limit in moduli space, that is the locus where the pre-potential is well approximated by Eq.~\eqref{Flcs}, we always have $|W_0|\gg 1$. This means that consistent heterotic moduli stabilisation in the large complex structure limit is unlikely, perhaps impossible. Motivated by this second no-go theorem we have explored the heterotic flux superpotential away from the large complex structure limit, using the full expressions for the pre-potential and the periods.\\[2mm]
This has been carried out for a number of examples, for which the period expressions are available in the literature, starting with the four one-parameter models obtained as mirrors of Fermat hypersurfaces in weighted projective space. A systematic scan over flux integers $(n_A,m^A)$ in the range defined by $|n_A|,|m^A|\leq 12$ for these four cases shows that the smallest possible $|W_0|$ values which can be obtained are $|W_0|={\cal O}(1)$. This result has been confirmed by a heuristic search, using genetic algorithms, in a larger range defined by $n_A,m^A\in\{-127,\ldots,128\}$. Values $|W_0|={\cal O}(1)$ may be considered marginally compatible with consistent moduli stabilisation.\\[2mm]
Finally, we have considered a two-parameter model and performed a systematic scan over fluxes $(n_A,m^A)$ in the range $|n_A|,|m^A|\leq 8$. In this case, the smallest values of $|W_0|$ are also found to be of order one. Thus, increasing the number of complex structure moduli did not lead to a reduction of the $|W_0|$ values.\\[2mm]
Our results show that small $|W_0|$ values for heterotic flux superpotentials are not easy to obtain. They are not available in the large complex structure limit, so that knowledge of the full period functions is required for further progress. The values $|W_0|={\cal O}(1)$ found in both the one- and two-parameter models indicate that achieving parametrically small flux superpotentials is challenging. At the same time, values of this magnitude may still be sufficiently small to allow for a non-trivial interplay with non-perturbative effects, and hence provide a viable starting point for consistent moduli stabilisation.\\[2mm]
There are many future directions worth pursuing, including extending our results to larger classes of CY manifolds, incorporating higher-order corrections, and attempting to stabilise all moduli. In particular, it would be interesting to extend the present analysis to known one-parameter and two-parameter models from the AESZ list~\cite{Almkvist:2005qoo} in order to study the distribution of $|W_0|$ values more systematically.

\section*{Acknowledgements}
EIB is supported in part by the Australian Research Council, project No. DP230101629.
AC is supported by the Royal Society grant DHF/R1/231142. AL is supported by the STFC consolidated grant ST/X000761/1. LL is supported by the Oxford-Croucher Scholarship for doctoral studies. We would also like to thank Thomas Harvey and Kit Fraser-Taliente for valuable discussions. EIB would like to thank Physics Department at Oxford University, where some of this work was done, for warm hospitality.
Finally, we thank the International Centre for Mathematical Sciences
(ICMS), Edinburgh, for hospitality during the final stages of writing this paper.\\[5mm] 
\appendix
\addcontentsline{toc}{section}{Appendix}
{\noindent\bf\Large Appendix}
\section{Proof of no-go theorem 2} \label{app:nogo2}
In this appendix we prove the no-go Theorem~\ref{thm:nogo-1} for supersymmetric vacua in the large complex structure limit. In this limit, the pre-potential $\mathcal{F}$ has already been presented in Eq.~\eqref{Flcs} and the resulting (affine versions of the) flux superpotential, $\hat{W}$, and K\"ahler potential, $\hat{K}$, in terms of the affine coordinates $Z^a=z^a+i\chi^a$, are given in Eqs.~\eqref{Wlcs} and \eqref{Klcs}. Recall, for these expression to be consistent we need the `mirror volume' $\tilde{\kappa}=\tilde{d}_{abc}z^az^bz^c$ to be large, that is, $\tilde{\kappa}\gg 1$.\\[2mm]
We introduce the notation
\begin{equation}
\tilde{\kappa}=\tilde{d}_{abc}z^az^bz^c\;,\quad \tilde{\kappa}_a=\tilde{d}_{abc}z^bz^c\; ,\quad \tilde{\kappa}_{ab}=\tilde{d}_{abc}z^c\; ,\quad \tilde{\kappa}' = \tilde{\kappa}+\frac{3k}{4}\; ,
\end{equation}
in terms of which the derivatives of the K\"ahler potential~\eqref{Klcs} can be written as
\begin{equation}
 \hat{K}_a:=\frac{\partial \hat{K}}{\partial Z^a}=-\frac{3\tilde{\kappa}_a}{2\tilde{\kappa}'}\;,\qquad
 \hat{K}_{ab}=\frac{\partial^2 \hat{K}}{\partial Z^a\partial \bar{Z}^b}=-\frac{3}{2}\left[\frac{\tilde{\kappa}_{ab}}{\tilde{\kappa}'}-\frac{3}{2}\frac{\tilde{\kappa}_a\tilde{\kappa}_b}{\tilde{\kappa}^{'2}}\right]\;.\label{Kab}
\end{equation}
It is useful to introduce the quantities $C=1+3k/(4\tilde{\kappa})$ and $C'=1-9k/(4\tilde{\kappa}')$, which are close to one in the limit $\tilde{\kappa}\gg 1$, and the covariant coordinates 
\begin{equation}
    z_a:=\hat{K}_{ab}z^b=\frac{3}{4}\frac{\tilde{\kappa}_a}{\tilde{\kappa}'}C' \qquad\Rightarrow\qquad z_az^a=\frac{3\tkap}{4\tkap'}C'\; .
\end{equation}
The F-term equations then take the form
\begin{equation} \label{Fa}
 F_a=\frac{\partial \hat{W}}{\partial Z^a}+\frac{\partial \hat{K}}{\partial Z^a}\hat{W}=\frac{\partial \hat{W}}{\partial Z^a}-\frac{2z_a}{C'}\hat{W}\stackrel{!}{=}0\; ,
\end{equation}
with the various parts explicitly given by  
\begin{eqnarray}
    {\rm Re}(\hat{W})&=&w\left[n_0'-n_a'\chi^a-\frac{1}{2}\tilde{d}_{abc}(m^a+m^o\chi^a)z^bz^c+\frac{1}{6}\tilde{d}_{abc}(3m^a+m^0\chi^a)\chi^b\chi^c\right]\label{ReW}\\
    {\rm Im}(\hat{W})&=&
    w\left[X+\frac{m^0}{6}(\tilde{\kappa}+3k)\right]\;,\qquad 
    X:=n_a'z^a-\frac{1}{2}\tilde{d}_{abc}(2m^a+m^0\chi^a)z^b\chi^c\label{ImW}\\
    {\rm Re}\left(\frac{\partial \hat{W}}{\partial Z^a}\right)&=&-w\left[\tilde{d}_{abc}\left(m^b+m^0\chi^b\right)z^c\right]\label{RedW}\\
    {\rm Im}\left(\frac{\partial \hat{W}}{\partial Z^a}\right)&=&w\left[n_a'-\frac{1}{2}\tilde{d}_{abc}\left(2m^b+m^0\chi^b\right)\chi^c+\frac{m^0}{2}\tilde{d}_{abc}z^bz^c\right]\label{ImdW}\;.
\end{eqnarray}
First consider the case $m^0 \neq 0$. Taking the imaginary parts of the F-terms and multiplying with $z^a$ gives
\begin{equation}
 0=z^a\,{\rm Im}(F_a)=z^a\,{\rm Im}\left(\frac{\partial \hat{W}}{\partial Z^a}\right)-\frac{3\tilde{\kappa}}{2\tilde{\kappa}'}{\rm Im}(\hat{W})=w\left[X+\frac{m^0}{2}\tilde{\kappa}\right] -\frac{3\tilde{\kappa}}{2\tilde{\kappa}'}{\rm Im}(\hat{W})
\end{equation}
We can use this equation to solve for $X$ and use the result to replace $X$ in the expression for ${\rm Im}(\hat{W})$ in Eq.~\eqref{ImW}. We find
\begin{equation}\label{ImW0local1}
{\rm Im}(\hat{W})=w\frac{2m^0}{3}\tilde{\kappa}\left(1+\frac{3k}{4\tilde{\kappa}}\right)\;,\qquad {\rm Im}(e^{\hat{K}/2}\hat{W})=w\left[\frac{m^0}{\sqrt{3}}\tilde{\kappa}^{1/2}\right] \left(1+\frac{3k}{4\tkap}\right)^{\frac{1}{2}}\;,
\end{equation}
and this is the result we wanted to prove for $m^0\neq 0$.\\[2mm]
It remains to discuss the case $m^0=0$. Evidently, in this case we have ${\rm Im}(\hat{W})=0$ from Eq.~\eqref{ImW0local1}, so there is no obstruction to a small $|W_0|$ from the imaginary part of $W$. But we still have to worry about ${\rm Re}(\hat{W})$. Specialising the F-term equations~\eqref{Fa} to $m^0=0$, using Eqs.~\eqref{RedW}, \eqref{ImdW} and \eqref{ReW}, we find
\begin{eqnarray}
 {\rm Re}(F_a)&=&{\rm Re}\left(\frac{\partial \hat{W}}{\partial Z_a}\right)-2z_a'\,{\rm Re}(\hat{W})=-w\tilde{\kappa}_{ab}m^b-2z_a'\,{\rm Re}(\hat{W})\stackrel{!}{=}0\label{RedWm0}\\
 {\rm Im}(F_a)&=&{\rm Im}\left(\frac{\partial \hat{W}}{\partial Z_a}\right)-2z_a'\,{\rm Im}(\hat{W})=w\left[n_a'-\tilde{d}_{abc}m^b\chi^c\right]\stackrel{!}{=}0\label{ImdWm0}
\end{eqnarray}
where we have used ${\rm Im}(\hat{W})=0$ in the second line and we have defined $z_a' = z_a/C'$. The imaginary parts of the F-term equations~\eqref{ImdWm0} can be written as 
\begin{equation}\label{chieq}
 M_{ab}\chi^b=n_a'\; , \qquad M_{ab}:=\tilde{d}_{abc}m^c\; .
\end{equation}
For any solution $\chi$ of this equation the real part of $\hat{W}$ becomes 
\begin{equation}\label{ReW0m0}
 {\rm Re}(\hat{W})=w\left[\tilde{n}_0-\frac{2\tkap'}{3C'}\,m^bz_b\right]\quad\mbox{where}\quad \tilde{n}_0
 :=n_0'-\frac{1}{2}n'_a\chi^a\; .
\end{equation}
Inserting this result into the real parts of the F-term equations~\eqref{RedWm0} gives, 
\begin{equation}\label{ReFa0}
 {\rm Re}(F_a)
 =\frac{2w}{3}\left[\tilde{\kappa}'\hat{K}_{ab}m^b-2\tkap'z_a'z_b'm^b-3\tilde{n}_0z_a'\right]\stackrel{!}{=}0\quad\;\Rightarrow\quad\;
\frac{2m^bz_bz^a}{{C'}^2}+3\tilde{n}_0\frac{z^a}{C'\tilde{\kappa}'}=m^a\; .
\end{equation}
The task is to solve the last equation for $z^a$ and two useful intermediate results are 
\begin{equation}\label{zsol0}
   m^bz_b=-\frac{9\tilde{n}_0}{2\tkap'}C_1\;,\quad  
   \tilde{n}_0z^a=-\frac{\tilde{\kappa}'}{6}m^a C_2\;,\quad
    C_1 := \frac{2\tkap(2\tkap-3k)}{(4\tkap+3k)(\tkap+3k)}\;, \quad C_2 := \frac{4(\tkap+3k)}{4\tkap+3k}\;.
\end{equation}
There are two cases here. If $\tilde{n}_0=0$ it follows  from the second Eq.~\eqref{zsol0} that all $m^a=0$ which, from Eq.~\eqref{chieq}, means that $M=0$ and, hence, that all $n_a'=0$. Further from Eq.~\eqref{ReW0m0} we see that $n_0'=0$. In short, $\tilde{n}_0=0$ implies that all fluxes vanish so this case is trivial and of no real interest. On the other hand, if $\tilde{n}_0\neq 0$ we can solve the second Eq.~\eqref{zsol0} to obtain solutions for $z^a$, namely
\begin{equation}\label{zeq2}
    z^a=-\frac{\tilde{\kappa}'}{6\tilde{n}_0}C_2m^a\qquad \Longrightarrow\qquad
    \tilde{\kappa}=-\frac{\tkap(m)}{(6\tilde{n}_0)^3}(\tkap')^3C_2^3\;,\qquad
   \tilde{n}_0 = -\frac{C_2}{6} \frac{\tkap(m)^{1/3}\tkap'}{\tkap^{1/3}}\; ,
\end{equation}
where $\tilde{\kappa}(m)=\tilde{d}_{abc}m^am^bm^c$.
Inserting this solution into the expression~\eqref{ReW0m0} for ${\rm Re}(\hat{W})$ gives
\begin{equation} \label{ReW0local2}
{\rm Re}(\hat{W})=-\frac{2w}{3}\left(1+\frac{3k}{4\tilde{\kappa}}\right)\tilde{\kappa}(m)^{1/3}\tilde{\kappa}^{2/3}\;,\quad
    {\rm Re}(e^{\hat{K}/2}\hat{W})=-\frac{w}{\sqrt{3}}\left(1+\frac{3k}{4\tkap}\right)^{1/2}\tilde{\kappa}(m)^{1/3}\tkap^{1/6} \;.
\end{equation}
This is what we needed to show in the case $m^0=0$, completing our proof.

\section{Period expressions} \label{sec:periods_full}
In this appendix we collect the period expansions of the manifolds explored in the present paper, which are the mirror Fermat families and the two-parameter model in $\mathbb{P}^{(1,1,2,2,2)}$. The detailed period expansions are taken from Refs.~\cite{Candelas:1990rm,Klemm:1992tx,Candelas:1993dm,Berglund:1993ax}.
\subsection{Mirror Fermat families}
Recall from Section~\ref{sec:one-param}, that the full basis of periods can be obtained from the fundamental period using $\varpi_j(\psi) = \varpi_0(\alpha^j \psi)$. For the mirror Fermat families listed in Table~\ref{tab:fermat_folds}, the fundamental period in the region $|\psi|>1$ is given by,
\begin{equation}
    \varpi_0(\psi) = \sum_{m=0}^\infty \frac{(km)!}{\prod_{i=0}^4(m\nu_i)!}(\gamma\psi)^{-km} \quad\mbox{where}\quad |\psi| > 1, \quad 0\leq \arg\psi<\frac{2\pi}{k}\; .
\end{equation}
After analytical continuation into the $|\psi|<1$ region, we can exploit Eq.~\eqref{eq:comp_basis} to write down a complete set of periods $\varpi_j(\psi)$, where $j=0,\ldots ,k-1$, given by
\begin{equation}\label{pijm}
    \varpi_j(\psi) = -\frac{\pi}{k} \sum_{n=1}^\infty \frac{1}{\Gamma(n)\prod_{i=0}^4\Gamma(1-n\nu_i/k)} \frac{e^{\frac{i\pi(k-1)n}{k}}}{\sin\left(\frac{\pi n }{k}\right)}(\gamma\alpha^j\psi)^n\;.
\end{equation}
Here, $\alpha = \exp\left(\frac{2\pi i}{k}\right)$, and the quantities $\gamma$, $n_i$ and $\nu_i$ have been defined in Table~\ref{tab:fermat_folds}. For convenience, we specialise the periods~\eqref{pijm} to the four cases $k=5,6,8,10$ by inserting the quantities $k$, $\alpha$ and $\gamma$ from Table~\ref{tab:fermat_folds}. This leads to 
\begin{align}
    \varpi_j(\psi)_{k=5} &= -\frac{1}{5} \sum_{m=1}^\infty \frac{\alpha_5^{2m}\Gamma(m/5)(5\psi)^m}{\Gamma(m)\Gamma^4(1-m/5)}\;,\\
    \varpi_j(\psi)_{k=6} &= -\frac{1}{6} \sum_{n=1}^\infty  \frac{\Gamma(n/6) (\gamma_6 \alpha_6^{j+5/2} \psi)^n}{\Gamma(n)\Gamma(1-n/3)\Gamma^3(1-n/6)}\;,\\
    \varpi_j(\psi)_{k=8} &= -\frac{1}{8} \sum_{n=1}^\infty  \frac{\Gamma(n/8) (\gamma_8 \alpha_8^{j+7/2} \psi)^n}{\Gamma(n)\Gamma(1-n/2)\Gamma^3(1-n/8)}\;,\\
    \varpi_j(\psi)_{k=10} &= -\frac{1}{10} \sum_{n=1}^\infty  \frac{\Gamma(n/10) (\gamma_{10} \alpha_{10}^{j+9/2} \psi)^n}{\Gamma(n)\Gamma(1-n/2)\Gamma(1-n/5)\Gamma^2(1-n/10)}\;,
\end{align}
where $j=0,\ldots, k-1$.
The analytical continuation to the region $|\psi| \geq 1$ leads to the periods~\cite{Klemm:1992tx},
\begin{equation} \label{eq:period_exp_out}
    \varpi_j(\psi) = -\frac{1}{(2\pi i)^3}\frac{1}{\prod_{i=0}^4\nu_i} \sum_{r=0}^3 \log^r(\gamma\psi) \sum_{N=0}^\infty \frac{(kN)!}{\prod_{i=0}^4(\nu_iN)!}b_{jrN}(\gamma\psi)^{-kN}\;,
\end{equation}
with the coefficients $b_{jrN}$ given by
\begin{align}
    b_{j0N} &= (2\pi i)^3 \left(S_{j4} - \prod_{i=0}^4\nu_i\right) + (2\pi i)^2 \left( 2\pi i + k\phi(N) \right) S_{j3}\nonumber\\
    &\quad+\frac{1}{24}(2\pi i)\left[ (2\pi i)^2 \left(4 + k^2 - \sum_{i=0}^4 \nu_i^2\right) + 12(2\pi i)k\phi(N) + 12 \left(k^2\phi^2(N) - k\phi'(N)\right) \right]S_{j2}\nonumber\\
    &\quad+\frac{k}{24}\left[ (2\pi i)^2 \left(k^2 - \sum_{i=0}^4 \nu_i^2\right)\phi(N) + 4k^2\phi^3(N) - 12 k\phi(N)\phi'(N) + 4\phi''(N)\right]S_{j1}\\
    b_{j1N} &= k(2\pi i)^2 S_{j3} + \frac{k}{2} (2 \pi i)\left( 2\pi i + 2k\phi(N) S_{j2} \right) \nonumber\\
    &\quad+\frac{k}{24}\left[ (2\pi i)^2 \left(k^2-\sum_{i=0}^4 \nu_i^2\right) + 12k^2\phi^2(N) - 12k\phi'(N) \right]S_{j1}\\
    b_{j2N} &= \frac{k^2}{2} \left( 2\pi i S_{j2} + k\phi(N)S_{j1} \right)\\
    b_{j3N} &= \frac{k^3}{6}S_{j1}\;.
\end{align}
Further, we have defined
\begin{equation}
    \phi(N) = \frac{1}{k} \sum_{i=0}^4 \nu_i \psi(1+\nu_i N) - \psi(1+kN)\;,
\end{equation}
\begin{equation}
    \psi(x) = \frac{d \log\Gamma(x)}{dx}\;,
\end{equation}
\begin{equation} \label{eq:s_fac_klemm}
    S_{jm} = \sum_{l=0}^{k-1} \alpha^{l(j+1)} \frac{\prod_{i=0}^4 (\alpha^{l\nu_i} -1)}{(\alpha^l-1)^{m+1}}\;.
\end{equation}
It can be explicitly checked that  Eq.~\eqref{eq:period_exp_out} for the $k=5$ case (the mirror quintic) reduces to the period expressions $\varpi_j(\psi)$ given in Appendix B of Ref.~\cite{Candelas:1990rm}.
As mentioned before, the periods $\varpi_j(\psi)$, for $j=0,\ldots ,k-1$ are not linearly-independent but are subject to $k-4$ linear relations. For the four cases, these linear relations are explicitly given by
\begin{equation}
\begin{array}{ccl}
    k=5&:& \varpi_0+\varpi_1+\varpi_2 + \varpi_3 +\varpi_4 = 0\\
    k=6&:& \varpi_0+\varpi_2+\varpi_4 =\varpi_1+ \varpi_3 +\varpi_5 = 0\\
    k=8&:& \varpi_j + \varpi_{j+4} = 0\;, \quad j=0,1,2,3\\
    k=10&:&\varpi_j + \varpi_{j+5} = 0\;, \quad j=0,1,2,3,4\\ && \varpi_0+\varpi_2 + \varpi_3 +\varpi_4+ \varpi_5 = 0
    \end{array}\; .
\end{equation}
Therefore, there are only four linearly-independent periods, which, for each of the above four cases, can be taken to be $(\varpi_0,\varpi_1,\varpi_2,\varpi_{k-1})$. As mentioned in the text around Eq.~\eqref{pidef} in the main text, this allows us to construct the symplectic period vector $\Pi$,
\begin{equation} \tag{\ref{pidef}}
    \Pi=\left(\begin{array}{l}\cF_0\\\cF_1\\\cZ^a\\\cZ^1\end{array}\right)=M_k\varpi\qquad\mbox{where}\qquad
    \varpi=-\frac{(2\pi i)^3}{\text{ord}(G)}\left(\begin{array}{l}\varpi_{2}\\\varpi_1\\\varpi_0\\\varpi_{k-1}\end{array}\right)\; ,
\end{equation}
and the `monodromy matrix' $M_k$ for each of the families is given by,
\begin{eqnarray} \label{eq:mmatrix}
    M_5 &= \left(\begin{array}{rrrr}
        -\frac{3}{5} & -\frac{1}{5} & \frac{21}{5} & \frac{8}{5}\\
        0 & 0 & -1 & 0\\
        -1 & 0 & 8 & 3\\
        0 & 1 & -1 & 0\\
    \end{array}\right)\;, \quad 
    &M_6 = \left(\begin{array}{rrrr}
        -\frac{1}{3} & -\frac{1}{3} & \frac{1}{3} & \frac{1}{3}\\
        0 & 0 & -1 & 0\\
        -1 & 0 & 3 & 2\\
        0 & 1 & -1 & 0\\
    \end{array}\right)\;, \nonumber \\[2mm]
    M_8 &= \left(\begin{array}{rrrr}
        -\frac{1}{2} & -\frac{1}{2} & \frac{1}{2} & \frac{1}{2}\\
        0 & 0 & -1 & 0\\
        -1 & 0 & 3 & 2\\
        0 & 1 & -1 & 0\\
    \end{array}\right)\;, \quad 
    &M_{10} = \left(\begin{array}{rrrr}
        0 & 1 & 1 & 1\\
        0 & 0 & -1 & 0\\
        1 & 0 & 0 & -1\\
        0 & 1 & -1 & 0\\
        \end{array}\right)\;.
\end{eqnarray}

\subsection{A two parameter model} \label{subsec:period_full_two}
The procedure for obtaining a full set of periods for the two parameter case follows very closely the analogous one parameter case. For our manifold in the weighted projective space $\mathbb{P}^{(1,1,2,2,2)}$, the fundamental period in the region $|8\psi^4/(\phi\pm1)|>1$ and $|\phi| > 1$ is,
\begin{equation}
    \varpi_0(\phi,\psi) = \sum_{r,s = 0}^\infty \frac{(8r+4s)!(-2\phi)^s}{((2r+s)!)^3(r!)^2s!(8\psi)^{8r+4s}}\;.
\end{equation}
We would like to find the period expansions in the region $|8\psi^4/(\phi\pm1)|<1$ and $|\phi|<1$. This requires two separate analytical continuations. We first analytically continue into the $|8\psi^4/(\phi\pm1)|<1$ region, and then exploit the relation $\varpi_j(\psi,\phi) = \varpi_0(\alpha^j\psi,\alpha^{4j}\phi)$ to find the full set of periods in the region $|8\psi^4/(\phi\pm1)|<1$ and $|\phi| > 1$. The result is
\begin{equation}
    \varpi_j(\psi,\phi) = -\frac{1}{4} \sum_{m=1}^\infty\frac{(-1)^m\alpha^{mj}\Gamma(\frac{m}{4})}{\Gamma(m)\Gamma^3\left(1-\frac{m}{4}\right)}(2^{12}\psi^4)^{\frac{m}{4}} u_{-\frac{m}{4}(\phi)}\;,
\end{equation}
where $j=0,\ldots ,7$ and the function $u_\nu(\phi)$ is defined in temrs of a the hypergeometric function as
\begin{equation}
    u_\nu(\phi) = (2\phi)^\nu \, {}_2F_1\left(-\frac{\nu}{2},\,-\frac{\nu}{2}+\frac{1}{2};\, 1;\,\frac{1}{\phi^2}\right)\;,
\end{equation}
and $\alpha = \exp\left(\frac{2\pi i}{8}\right)$. Now we further analytically continue this function $u_\nu(\phi)$ from $|\phi|>1$ to $|\phi| < 1$ where it is described by the series
\begin{equation}
    u_\nu(\phi) = \frac{e^{i\pi\nu/2}}{2\Gamma(-\nu)} \sum_{n=0}^\infty \frac{e^{i\pi n /2} \Gamma\left(\frac{n-\nu}{2}\right) (2\phi)^n}{n! \Gamma\left(1-\frac{n-\nu}{2}\right)}\; .
\end{equation}
This leads to the full set of periods in the region  $|8\psi^4/(\phi\pm1)|<1$ and $|\phi|<1$, given by
\begin{equation}
    \varpi_j(\psi,\phi) = -\frac{1}{4}\sum_{m=1}^\infty\sum_{n=0}^\infty C_{mnj} \psi^m\phi^n\;,
\end{equation}
where $j=0,\ldots ,7$ and  the coefficients $C_{mnj}$ are defined as
\begin{equation}
    C_{mnj} = \frac{(-1)^{m+jn}\cdot2^{3m+n-1}\cdot e^{\frac{i\pi}{8}(2jm-m+4n)}}{n!} \frac{\Gamma\left(\frac{n}{2}+\frac{m}{8}\right)}{\Gamma(m)\Gamma^3\left(1-\frac{m}{4}\right)\Gamma\left(1-\frac{n}{2}-\frac{m}{8}\right)}\;.
\end{equation}
These periods are linearly dependent and are subject to the two linear relations,
\begin{equation}
    \sum_{j=0}^3\varpi_{2j}(\psi,\phi) =  \sum_{j=0}^3\varpi_{2j+1}(\psi,\phi)=0\;.
\end{equation}
There are therefore six linearly independent periods which we can choose to be $(\varpi_0,\varpi_1,\varpi_2,\varpi_3,\varpi_4,\varpi_5)$. There six periods together with the `monodromy matrix' $M$ determine the period vector $\Pi$. It is explicitly given by 
\begin{equation} \tag{\ref{eq:period_2_defn}}
    \Pi(\psi) = \begin{pmatrix}
        \cF_0\\ \cF_1\\\cF_2\\ \cZ^0\\ \cZ^1\\\cZ^2
    \end{pmatrix}= M \varpi(\psi) \;,
    \quad \text{where}\quad
    \varpi = 
    \begin{pmatrix}
        \varpi_0\\ \varpi_1\\ \varpi_2\\ \varpi_3\\\varpi_4\\ \varpi_5
    \end{pmatrix}\;,
\end{equation}
where the $6\times 6$ matrix $M$ is given by
\begin{equation}
\renewcommand{\arraystretch}{1.4}
    M = \left(\begin{array}{rrrrrr}
    -1 & 1 & 0 & 0 & 0 & 0 \\
    1 & 0 & 1 & -1 & 0 & -1 \\
    \frac{3}{2} & 0 & 0 & 0 & -\frac{1}{2} & 0 \\
    1 & 0 & 0 & 0 & 0 & 0 \\
    -\frac{1}{4} & 0 & \frac{1}{2} & 0 & \frac{1}{4} & 0 \\
    \frac{1}{4} & \frac{3}{4} & -\frac{1}{2} & \frac{1}{2} & -\frac{1}{4} & \frac{1}{4}
    \end{array}\right).
\end{equation}

\bibliography{main}
\bibliographystyle{ieeetr}

\end{document}